\documentclass[12pt]{emulateapj}

\def \bea {\begin{eqnarray}}
\def \ena {\end{eqnarray}}                  
\def \bee {\begin{equation}}
\def \ene {\end{equation}}
\def    \bJ     {\bf J}
\def    \ba     {\bf  a}

\def    \bB     {\bf  B}

\def \mc {\mbox{cos }}
\def \ms {\mbox{sin }}

\begin{document}
\shorttitle{Radiative torques alignment}
\shortauthors{Hoang \& Lazarian}
\title{Radiative torques alignment in the presence of pinwheel torques }

\author{Thiem Hoang \& A. Lazarian}
\affil{Dept. of Astronomy, University of Wisconsin,
   Madison, WI53706; hoang; lazarian@astro.wisc.edu}

\begin{abstract}
We study the alignment of grains subject to both radiative torques and pinwheel torques while accounting for thermal flipping of grains. By pinwheel torques we refer to all systematic torques that are fixed in grain body axes, including the radiative torques arising from scattering and absorption of isotropic radiation. We discuss new types of pinwheel torques, which are systematic torques arising from infrared emission and torques arising from the interaction of grains with ions and electrons in hot plasma. We show that both types of torques are long-lived, i.e. may exist longer than gaseous damping time. We compare these torques with the torques introduced by E. Purcell, namely, torques due to H$_2$ formation,
the variation of accommodation coefficient for gaseous collisions and photoelectric emission. 
Furthermore, we revise the Lazarian \& Draine model for grain thermal flipping. We calculate mean flipping timescale induced by Barnett and nuclear relaxation for both paramagnetic and superparamagnetic grains, in the presence of stochastic torques associated  with pinwheel torques, e.g. the stochastic torques arising from H$_2$ formation, and gas bombardment. We show that the combined effect of internal relaxation and stochastic torques can result in fast flipping for sufficiently small grains and, because of this, they get thermally trapped, i.e. rotate thermally in spite of the presence of pinwheel torques. For sufficiently
large grains, we show that the pinwheel torques can increase the degree of grain alignment achievable with the radiative torques by increasing the magnitude of the angular momentum of low attractor points and/or by driving grains to new high attractor points.
\end{abstract}


\section{Introduction}
Polarization of radiation arising from emission or absorption by aligned grains is widely used to study magnetic fields in the diffuse interstellar medium (see Goodman et al. 1995), molecular clouds (see Hildebrand et al. 2000, 2002) and in prestellar cores (Crutcher et al. 2004). The grain alignment involves the alignment of the axis of major inertia $\ba_{1}$ with angular momentum $\bJ$ and the alignment of $\bJ$ with respect to the ambient magnetic field. 

An understanding of grain dynamics is extremely important for both understanding of dust properties and
dust alignment. The latter is essential for determining situations when polarization of starlight passing through dusty interstellar gas, as well as polarization of dust emission can be
reliably interpreted in terms of magnetic field direction. In terms of dust properties, fast rotation should
disrupt loose aggregates, placing limits for the fractal dimensions of dust particles. 

As dust scatters or absorbs photons, it experiences radiative torques. These torques can be stochastic
or systematic. Stochastic radiative torques arise from, for instance, a spheroidal grain randomly emitting or
absorbing photons. The latter process, for instance, was invoked by Harwit (1970) in his model of 
grain alignment based on grains being preferentially spun up in the direction perpendicular to the 
photon beam. However, Purcell \& Spitzer (1971) showed that randomization arising from the same
grain emitting thermal photons makes the achievable degree of grain alignment negligible. In fact,
they showed that random radiative torques arising from grain thermal emission is an important 
process of grain randomization. More recently, grain thermal emission was analyzed as a source
of the excitation of grain rotation as well as its damping in relation to the rotation of tiny spinning
grains that are likely to be responsible for the so-called anomalous foreground emission 
(Draine \& Lazarian 1998). 

Systematic radiative torques were first introduced by Dolginov (1972) in terms of chiral, e.g. hypothetical quartz grains.
Starlight passing through such grains would spin them up. Later, Dolginov \& Mytrophanov (1976)
considered an irregular grain model, consisting of two twisted spheroids, made of more accepted materials, e.g. silicate grains, and claimed
that these grains will be both spun up and aligned by radiative torques arising from the anisotropic component of radiation field.\footnote{Using
so-called  Rayleigh-Gans approximation, Dolginov \& Silantiev (1976) provided calculations of the radiative torques for a model of an irregular grain consisting of two ellipsoids twisted with respect to each other, but our calculations of radiative torques using the Discrete Dipole Approximation code (Draine \& Flatau 2004) for
the same set parameters and the same model as in the aforementioned work are in conflict with their
analytical findings (see more in Hoang \& Lazarian 2008b).} Further on, following the convention we adopted in Lazarian \& Hoang (2007a, hereafter LH07a), we shall call these torques RATs. The work by Dolginov \& Mytraphanov (1976) was, unfortunately, mostly ignored for 20 years. A possible explanation of
this may
be due to the fact, that for a long time, the magnitude of RATs for realistic irregular grains
remained unclear. 

A renewed interest to RATs was induced by the possibility of calculating them for arbitrary grain shapes. This occurred after Bruce Draine modified correspondingly
his publicly-available Discrete Dipole Approximation code (hereafter DDSCAT; Draine \& Flatau 2004). Moreover, Draine \& Weingartner (1996, 1997, henceforth DW96, DW97, respectively) conjectured that 
RATs may provide the primary alignment mechanism for interstellar grains. Support for this claim came through later research in the field, by better understanding the dynamics
of grains subject to RATs from anisotropic radiation fields, calculating grain alignment with an
analytical model (LH07a), as well as including important physical processes like 
grain wobbling (Lazarian 1994; Lazarian \& Roberge 1997; Weingartner \& Draine 2003; Hoang \& Lazarian
2008a), and accounting within the model for gaseous bombardment and uncompensated
torques arising from H$_2$ formation, as was done in Hoang \& Lazarian (2008a, henceforth HL08a). As it was described
in a review by Lazarian (2007), RAT alignment mechanism has become not only the leading
candidate to explain interstellar grain alignment, but also to explain grain alignment in many other astrophysical
environments, including circumstellar regions (Aitken et al. 2002), accretion disks (Cho \& Lazarian 2007), comet atmospheres
 (see Rosenbush et al 2007), and molecular clouds (Whittet et al. 2008).

 In LH07a, we subjected to scrutiny the properties of RATs.
Using a simple analytical model (AMO) of a helical grain we studied the properties
of RATs and the RAT alignment. The results obtained by the AMO were shown to 
be in good correspondence with numerical calculations obtained by DDSCAT for irregular grains. Invoking 
the generic properties of the RAT components, we explained
the RAT alignment of grains in both the absence and presence of magnetic fields. 
Intentionally, for the sake of simplicity, in LH07a
 we studied a simplified dynamical model to demonstrate the effect of RATs. Within the latter model we showed that RATs can align grains at attractor points with low angular momentum (i.e., $J$ of the order of thermal angular momentum $J_{th}$, hereafter, low-$J$ attractor points), and/or attractor points with high angular momentum (i.e., $J\gg J_{th}$, hereafter high-$J$ attractor points). The high-J attractor points mostly correspond to a perfect alignment of $\bJ$ with respect to $\bB$, while the low-J attractor points occur at perfect alignment angle or at some angle in the vicinity of the perfect alignment.

One of the effects that was not discussed in LH07a was the effect of thermal fluctuations.  Lazarian (1994) noticed that in spite of the fast rates of internal relaxation, rotating grains wobble. This
conclusion was a consequence of the Fluctuation-Dissipation Theorem (see Landau \& Lifshitz 1976), which states that any dissipation process, e.g. internal relaxation, should be accompanied by the proportionally
fluctuation process. It is possible to show that grains rotating thermally are expected to wobble with larger amplitude compared to their counterparts rotating at suprathermal (much greater than thermal) rates (see
Lazarian 1994, Lazarian \& Roberge 1997).
Therefore the zero value of angular momentum at low-J attractor points obtained in LH07a stemmed from the assumption of the perfect coupling of grain axis of the maximal moment of inertia
$\ba_{1}$ with $\bJ$, which
is violated for sufficiently low $J$. In fact, in HL08a, we treated fully the dynamics of the grain alignment by taking into account thermal fluctuations (see also Weingartner \& Draine 2003), and showed that thermal fluctuations within irregular grains can increase the angular momentum of the low-J attractor points from $J=0$ to $J\sim J_{d}$, i.e. the angular momentum of grains corresponding to dust grain temperature. In addition, HL08a showed that gas bombardment and other randomizing torques can contribute significantly to the RAT alignment by moving grains
from low-$J$ to high-$J$ attractor points, whenever the latter are present. However, we feel that a comprehensive study on the degree of alignment as a function of grain size and radiation intensity is very essential for polarization modeling. Note that earlier papers which attempted to introduce grain alignment into polarization modeling (Cho \& Lazarian 2005; Pelkonen et al. 2007; Bethell et al. 2007) assumed  perfect alignment for grains larger than some critical size, and no alignment for smaller grains. The latter was inferred using the criterion of whether RATs for radiation field under interest, which are  sharp function of grain size, can induce grains to rotate several times faster than the thermal rotation rate. This is a rather crude criterion because it does not take into account the type of RAT alignment, i.e with or without high-J attractor points. For instance, if RATs align all grains at low-$J$ attractor points, then  assumption above leads to overestimates for the degree of alignment. 

In addition to RATs, other uncompensated torques act on grains (Purcell 1979; LD99a; Roberge \& Ford 2000, henceforth RF00). Purcell (1979) proposed three processes that can produce uncompensated torques: hydrogen formation, photoelectric effect and the variation of accommodation coefficient. These pinwheel torques together with the paramagnetic dissipation were thought to be the major mechanism leading to the alignment of $\bJ$ with $\bB$. However, the paramagnetic dissipation was found to be unable to explain the alignment of paramagnetic grains in the weak interstellar magnetic field. In any case, in the presence of RATs, the alignment arising from paramagnetic dissipation
for ordinary paramagnetic grains is negligible compared to that arising from RATs\footnote{Unfortunately, the misconception that RATs act as proxies of Purcell's torques and the alignment arises from paramagnetic dissipation is well entrenched in
the grain alignment literature. This claim is based on the DW96 study and, in spite of the fact, that the efficient RAT alignment was demonstrated already in DW97 and later works, the claim persisted. This may be
due to the fact that, DW97 did not provide, unlike DW96, the predictions for the degree of alignment. The problem of predicting the degree of alignment happened to be a tough one. The present paper is an attempt in this direction. Interestingly enough, Dolginov \& Mytrophanov (1976) did consider RATs as a mechanism capable of aligning grains irrespectively of the presence or absence of paramagnetic relaxation. For superparamagnetic grains with sufficient number of iron atom per cluster, Lazarian \& Hoang (2008) showed that the alignment always happen with high-$J$ attractor points and is perfect. However, even in this case, RATs are more important than the enhanced paramagnetic relaxation. The latter effect causes the stabilization of the high-$J$ attractor point created by RATs.}.

The efficiencies of the Purcell torques decrease if grains wobble thermally as we discussed above. The effect of thermal wobbling was quantified in Lazarian \& Roberge (1997) and the
results of this study were used by Lazarian \& Draine (1999ab, henceforth LD99ab) to predict new effects of grain dynamics, namely, thermal flipping and thermal trapping. The thermal flipping is the effect
of occasional increase of the wobbling angle beyond 90 degrees. When this occurs, the torques that are fixed in grain body coordinates change their direction. In the case thermal flipping occurs
fast enough, the Purcell torques get averaged out and the grain rotates thermally in spite of the presence of the uncompensated pinwheel torques, i.e it is thermally trapped. 

However, this picture was challenged recently by Weingartner (2008). His study has several new mathematical points. The most important one is that he used a dimensionless variable $q=2I_{1}E/J^{2}$ with $I_{1}$ being the inertia moment along the axis of major inertia $\ba_{1}$ and $E$ being the total rotational energy to describe the internal relaxation, instead of the angle $\theta$ between $\ba_{1}$ and $\bJ$. He suggested an integration constant for diffusion coefficient of internal relaxation, which differs from that in the Lazarian \& Roberge (1997). We agree with the choice of the integration constant. For this choice in Weingartner (2008) the diffusion coefficient vanishes when  $\ba_{1}$ perpendicular to angular momentum $\bJ$. As a result, he found that grains do not experience thermal flipping as a result of internal relaxation. We agree with this choice of the integration constant as it corresponds to
the
nature of the Fluctuation-Dissipation Theorem employed in Lazarian (1994) to
describe the wobbling of the grains. Indeed, according to the theorem no
dissipation should correspond to no fluctuations. For an oblate grain the
internal dissipation goes to zero as $\theta \rightarrow \pi/2$. Thus the
Fluctuation-Dissipation Theorem suggests that the at this point the
fluctuations should also go to zero. One can check that the choice of the
constant in Weingartner (2008), which was obtained from more formal
considerations, corresponds to this physical requirement. However, as it
clear from the rest of the paper, we disagree that this modification of the
diffusion coefficients will preclude physical processes of thermal flipping
and trapping introduced in Lazarian \& Draine (1999) from happening. 

In what follows, we consider a more realistic picture of suprathermal rotation, namely, we take into account  the fact that any pinwheel torque is accompanied by stochastic torques of the same nature. For instance,
apart from systematic torques, the process of H$_2$ formation induces stochastic torques. In this model it is clear that the diffusion is present even when the angle between ${\bf a_1}$ and ${\bf J}$
is 90 degrees, which means that the thermal flipping is possible. In this paper we revisit the LD99a study and confirm that the thermal flipping and thermal trapping predicted in LD99a are important effects of grain dynamics.

The structure of the present paper is as followings. In \S 2 we present pinwheel torques, including long-lived and short-lived torques. We calculate the maximal value of angular momentum that these pinwheel torques can achieve for the ISM. In \S 3 we revise the problem of thermal flipping and calculate the thermal mean flipping time induced by internal relaxation in the presence of external stochastic torques. Critical size of flipping grain and trapping size for the ISM are estimated in this section. We study the effects of H$_{2}$ pinwheel torques and thermal flipping by stochastic torques accompanied with H$_{2}$ formation on the RAT alignment for the diffuse ISM in \S 4. We also calculate the value of $J$ at low attractor points as a function of the magnitude of H$_{2}$ pinwheel torques for different grain sizes in this section. We summarize our findings in \S 5.

\section{Pinwheel torques: Revisiting Purcell 1979}
Pinwheel torques were first discussed by Purcell (1979), who identified H$_2$ torques as the dominant pinwheel torques acting in interstellar gas.  With the existing
uncertainties in calculation of the torques this may or may not be true. Moreover, for other environments (e.g. molecular clouds), we know for certain that it is not true.
This induces our extended discussion of the pinwheel torques. Purcell also estimated that the $H_2$ torques should be short-lived and this induced further
researchers (see Spitzer \& McGlynn 1979; Lazarian 1995; Lazarian \& Draine 1997) to concentrate on the studies of short-lived pinwheel torques. 

Usually, we represent the magnitude of pinwheel torques in terms of the thermal angular momentum, and the gas damping time. For simplicity, we consider a spheroid grain, with the moments of inertia $I_{\|}=I_{1}, I_{\perp}=I_{2}=I_{3}$.\footnote{For oblate spheroid with major and minor axes a and b, $I_{\|}=2Ma^{2}/5=(8\pi\rho/15)ba^{4}$, and $I_{\perp}=(4\pi\rho/15) a^{2}b(a^{2}+b^{2})$, but sometimes we define the equivalent sphere with the same volume of the grain, and use $I_{1}=(8\pi/15)\rho a^{5}$.} They are defined by
\bea
J_{th}&=&\sqrt{I_{\|}k_{B}T_{gas}}=\sqrt{\frac{8\pi\rho a^{5}s}{15}k_{B} T_{gas}},\nonumber\\
&=&5.89\times 10^{-20} a_{-5}^{5/2}\hat{s}^{1/2}\hat{\rho}^{1/2} \hat{T}_{gas}^{1/2} \mbox{  g cm}^{2}\mbox{rad s}^{-1},\label{jtherm}
\ena
with $\rho$ density of material within the grain and $\hat{\rho}=\rho/3{~\mbox g cm}^{-3}$, $a$ grain size, $a_{-5}=a/10^{-5}{~\mbox cm}$, $\hat{s}=s/0.5$ with $s=b/a$ and $T_{gas}$ gas temperature and $\hat{T}_{gas}=T_{gas}/100$K. And the damping time due to gas collision (see Roberge, DeGraff \& Flatherty 1993, hereafter RDF93) is 
\bea
t_{gas}&=&\frac{3}{4\sqrt{\pi}}\frac{I_{\|}}{n_{H}m_{H} a^{4}v_{th}\Gamma_{||}},\nonumber\\
&=&2.3\times 10^{12} a_{-5}\hat{s}\hat{T}_{gas}^{-1/2}\left(\frac{30 \mbox{ cm}^{-3}}{n_{H}}\right) {~\mbox s},\label{tgas}
\ena
where $v_{th}$ is the thermal velocity of hydrogen atom with density $n_{H}$, $m_{H}$ is the hydrogen mass, $\Gamma_{\|}$ is the eccentricity parameter which is unity for sphere (see RDF93), and we adopted standard parameters in the ISM. 

We also define dimensionless angular momentum and torque
\bea
J'&=&\frac{J}{J_{th}},\label{jthermp},\\
\Gamma'&=&\Gamma \frac{t_{gas}}{J_{th}},\label{gammap}
\ena
where $J_{th}$  and $t_{gas}$ are given by equations (\ref{jtherm}) and (\ref{tgas}).

\subsection{Long-lived pinwheel torques}

\subsubsection{Radiative torques due to absorption and scattering of isotropic starlight}

While Dolginov \& Mytrophanov (1976) considered only torques arising from anisotropic radiative flows, DW96 noticed that radiative torques are not zero even if the
radiation field is isotropic. Such "isotropic radiative" torques have all the properties of the pinwheel torques introduced by Purcell (1979), but with the exception that these
torques were definitely long-lived. Indeed, to change them, one required to change the entire grain shape, rather than to change properties of grain surface.
The former, in most cases required time-scale longer than the gaseous damping time $t_{gas}$.

The isotropic radiative torques for studied grain shapes are usually a factor of $10^2$ weaker than the torques arising from anisotropic radiation. Therefore, in some subsequent
publications (e.g. in Cho \& Lazarian 2005, 2007; Pelkonen et al. 2007), they were disregarded. They may be occasionally important for grain spin-up. In this paper, we shall consider isotropic radiative torques as an example of long-lived pinwheel torques.

The isotropic radiative torque is defined by
\bea
Q_{iso}&=\frac{1}{4\pi}\int_{0}^{2\pi}\int_{0}^{\pi}Q_{a_{1}}(\Theta,\beta)\ms\Theta d\beta d\Theta,\label{qiso}
\ena
where $\Theta$ is the angle between the axis of major inertia $\ba_{1}$ and the radiation direction ${\bf k}$, $\beta$ is the rotation angle of the grain along $\ba_{1}$ (see Fig. \ref{amo}), and $Q_{a_{1}}(\Theta,\beta)={\bf Q}_{\Gamma}.\ba_{1}$ with ${\bf Q}_{\Gamma}$ being the radiative torque efficiency vector (see Appendix C) is the component of RATs along $\ba_{1}$. Here $Q_{a_{1}}(\Theta,\beta)$ does not depend on the precession angle $\Phi$ of the grain axis $\ba_{1}$ about ${\bf k}$.

\subsubsection{Radiative torques arising from grain infrared emission}

All systematic\footnote{Harwit (1970), Purcell \& Spitzer (1971) considered stochastic torques arising from emission of photons (see a more detailed discussion in Draine \& Lazarian 1999).} radiative torques so far were assumed to arise from absorption and scattering  of radiation by grain (Dolginov \& Mytrophanov 1976, DW96). However, it is possible
to show that pinwheel torques should also arise as a result of grain infrared emission. 

 To calculate the radiative torques arising from grain infrared emission, let us consider a grain at thermal equilibrium. In such conditions
 the grain should not be subject to systematic spin-up. This means that the torques arising from
the infrared emission $\Gamma_{em}$ should compensate the torques $\Gamma_{ext}$ arising from absorption and scattering of radiation, i.e.
\begin{equation}
\Gamma_{em}+\Gamma_{ext}=0.
\label{equil}
\end{equation}
Therefore, when $\Gamma_{ext}$ is known, equation~(\ref{equil}) allows us to find the torques
arising from grain emission. 
The pinwheel torque due to infrared emission is then
\bea
\Gamma_{em}=-\left(\frac{a^{2}{\bar{\lambda}(T_{d})} {u}_{rad}}{2}\right)\overline{Q_{iso}(T_{d})},\label{emis2}
\ena
where $a$ is the grain size, $\bar{\lambda}(T_{d})$ and ${u}_{rad}$ are the mean wavelength and energy density of infrared emission from the dust with temperature $T_{d}$, and $\overline{Q_{iso}}$ is the isotropic torque (see Appendix C) averaged over the spectrum of grain infrared emission. Here the infrared emission spectrum of dust grain with temperature $T_{d}$ is assumed to be the blackbody with the energy density per unit of wavelength $u_{\lambda}$:
\bea
u_{\lambda}=\frac{B_{\lambda}(T_{d})}{c},
\ena
where $B_{\lambda}(T_{d})$ is the brightness of the black body with temperature $T_{d}$ and $c$ is the light speed. 

We can calculate the maximum value of angular momentum induced by $\Gamma_{em}$  as
\bea
{J}_{max}^{'em}&=&\Gamma'_{em}=2.4\times10^{3}\left(\frac{\overline{\lambda}(T_{d})}{1.2 ~\mu m}\right)\left(\frac{{u}_{rad}}{u_{ISRF}}\right)\hat{\rho}^{1/2}a_{-5}^{1/2}\nonumber\\
&&\times \left(\frac{3000\mbox{ cm}^{-3}\mbox{K}}{n_{H}T_{gas}}\right)\overline{Q_{iso}(T_{d})},\label{jem}
\ena
where $u_{ISRF}$ is the mean energy density of the interstellar radiation field (hereafter, the ISRF; see Table 1).

\begin{figure}
\includegraphics[width=0.5\textwidth]{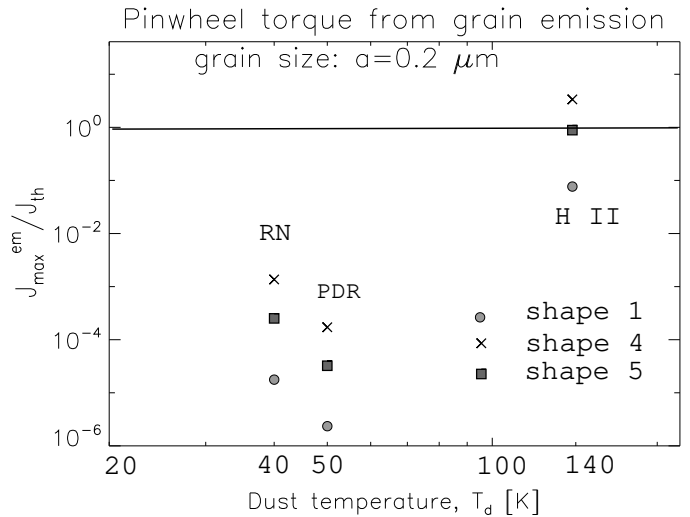}
\caption{$J_{max}^{em}/J_{th}$ excited by pinwheel torques due to infrared emission for a grain size $a=0.2 \mu m$ in reflection nebulae (RN), photodissociation region (PDR) and HII regions with dust temperature $T_{d}=40, 50$   $140$ K, gas temperature $T_{gas}=100, 300$ and $8\times 10^{3}$K, gas density $n_{gas}=10^{3}, 10^{5}$ and $5\times 10^{3}$cm$^{-3}$, respectively. Astronomical silicate grains of shape 1, 4 and 5 in LH07a are adopted.}
\label{qemis}
\end{figure}

To obtain $\overline{Q_{iso}(T_{d})}$, we first use DDSCAT (Draine \& Flatau 2004) to calculate RAT components $Q_{e1}, Q_{e2}$ and $Q_{e3}$, and then $Q_{iso}$ for different angles $\Theta$ ranging from $0$ to $\pi$ and $\beta$ in the range $0$ to $2\pi$ at $\Phi=0$ and for the entire spectrum of radiation field with $\lambda$ from $0.1\mu$m to $20\mu$m, for astronomical silicate grains of irregular shapes and of different sizes. Then, we average over the angles $\beta$ and $\Theta$ following equation (\ref{qiso}) to get ${Q_{iso}}$. Finally, we average ${Q_{iso}}$ over the entire spectrum of grain emission to obtain $\overline{Q_{iso}(T_{d})}$. Combining with equation (\ref{jem}) we obtain $J_{max}^{'em}$. Figure \ref{qemis} shows $J_{max}^{em}/J_{th}$ for grain shapes 1, 4 and 5 in LH07a  with the size $a=0.2\mu m$, embedded in reflection nebulae (RN), photodissociation (PDR) and HII regions with dust temperature $T_{d}=40, 50$ and $140$ K, gas temperature $T_{gas}=100, 300$ and $8\times 10^{3}$K, gas density $n_{gas}=10^{3}, 10^{5}$ and $5\times 10^{3}$cm$^{-3}$, respectively.

Figure \ref{qemis} shows that the infrared emission torques are of marginal importance for RN and PDR when $T_{d}$ is low. However, in H II regions, where the dust temperature can achieve $\sim 140$K as a result of the heating from both the radiation of O and B stars, $0.2\mu$m grain can rotate suprathermally, i.e., with $J\ge J_{th}$ (see Fig. \ref{qemis} for shape 4 and 5). Since radiative torques increases fast with the grain size (CL07, HL07a), larger grains are expected to rotate faster by infrared pinwheel torques. Therefore, we believe that this important source of
pinwheel torques should be included in modeling of grain alignment.

\subsubsection{Regular Torques from Grain-Electron interactions in plasma}

Heating of grains in plasma is well described in the literature (see Dwek et al. 2007). Electrons hit grains more frequently than ions. This effect is partially compensated by grains
getting negative charge. In situations when grain heating is mostly by electrons, the heating rate
depends on whether grains are large enough for electrons to stop within grain material or electrons pass
through grains. Below, we discuss, however, a different effect.

As electrons pass through grain material they exert torques. These torques can have random and systematic
components. Similarly, as for radiative torques, for homogeneous grains with symmetric shapes, e.g.
ellipsoids, prisms, systematic torques vanish. However, if grains are inhomogeneous or of irregular shapes,
we expect to observe Purcell-type torques arising from grain-plasma interactions. For instance, electrons
penetrating from different sites of a grain having a shape of prisms, but made of inhomogeneous material
may have different stopping distance, deflection rates or probabilities of ejecting secondary electrons.
Similar as in the Purcell (1979) study, these variations at different sites are difficult to quantify with a precise model. 
Therefore, as in Purcell (1979) one may introduce variations of grain properties using plausibility arguments. One quantitative estimate is beyond the scope of this paper.

\subsubsection{Regular Torques due to gas-grain interaction}

Below in \S 2.2.2 we discuss systematic torques arising from the variations of the accommodation coefficient. However, gaseous bombardment can produce torques related to grain shape. In Lazarian \& Hoang (2007b) we discussed torques arising from grain streaming\footnote{Unlike the traditional mechanical alignment processes (Gold
1952b, Purcell 1969, Purcell \& Spitzer 1971, Dolginov \& Mytrophanov 1976, Lazarian 1994, 1995b, 1997, Lazarian \& Efroimsky 1996, Efroimsky 2003, Roberge et al. 1995),
this process is efficient even if the gas-grain drift is subsonic.}, which we assumed helical and approximated with AMO. The torques, arising from such a gas-grain interaction
are analogous to RATs. It is only natural to assume the existence of the analog of the isotropic torques for grain-gas interactions. In terms of AMO this would
correspond to the opposite surfaces of the mirror interacting differently with the impinging atoms, i.e. very similar to the modification of AMO for representing of
isotropic radiative torques that we discussed in LH07a. Such torques should be long-lived, as the changes of the grain shape would be involved.

\subsection{Short-lived pinwheel torques}

Below, following Purcell (1979) we identify the torques due to H$_2$ formation and the variation of the accommodation coefficient over grain surface as short-lived pinwheel torques. It was argued in Lazarian (1995) that desorption of atoms can make these torques long-lived. Nevertheless, conservatively, we follow Purcell's approach. 

\subsubsection{Regular torques arising from H$_2$ formation}
H$_{2}$ formation followed by the ejection of H$_{2}$ molecules results in the systematic torque along the preferred rotation axis (i.e., axis of major inertia $\ba_{1}$). In addition, the H$_{2}$ formation on different active sites is random, and leads to the second order effect of torques. 

The detail calculations for the systematic torque component for a brick and a spheroid were presented in Purcell (1979) and Lazarian \& Roberge (1997), respectively. For a brick grain of sides $2a\times 2a\times 2b$ with $s=b/a<1$, the H$_{2}$ pinwheel torque in dimensionless units is given by
\bea
\langle \left(\Gamma'_{H_{2}}\right)^{2}\rangle^{1/2}&\approx&3.5\times 10^{3} (1-y)\hat{\gamma}_{H}\hat{s}^{5/2}a_{-5}^{1/2}\hat{\rho}^{1/2}\hat{T}_{gas}^{-1/2}\nonumber\\
&&\times \hat{E}_{kin}^{1/2}\left(\frac{10^{11}\mbox{cm}^{-2}}{\alpha}\right)^{1/2},\label{eq12}
\ena
where $y=2n(H_{2})/n_{H}$ being the fraction of hydrogen in the molecular state, $\hat{\gamma}_{H}=\gamma_{H}/0.1$ with $\gamma_{H}$ being the fraction of H atom converted to H$_{2}$ molecule, $\hat{s}=s/0.5$, $\alpha$ is the density of active site on the grain surface and $\hat{E}_{kin}=E_{kin}/0.2\mbox{eV}$ with $E_{kin}$ being the kinetic energy of H$_{2}$ molecule escaping from the grain surface. We can now write the maximal value of grain angular momentum achieved by H$_{2}$ pinwheel torque as
\bea
J_{max}^{'H_{2}}=\langle \left(\Gamma'_{H_{2}}\right)^{2}\rangle^{1/2}.\label{jh2max}
\ena

Equation (\ref{eq12}) shows that the H$_{2}$ pinwheel torque depends on the active site density $\alpha$. This parameter is not well constrained at the moment. Lazarian (1995) argued that  poisoning 
of active sites is more important than resurfacing and estimated that small grains with $a<10^{-5}$~cm should have active sites poisoned.

\subsubsection{Regular torques due to the variation of accommodation coefficient}

Purcell (1979) noticed that if gas and grain temperatures are different, the variation of the properties of the dust grain should create
systematic torques. For the variation of the accommodation coefficient of 10 percent with the scale of $1/3$ of the grain size Purcell (1979) obtained that for typical conditions of the diffuse medium the torques will be of the order $10^{-2}$ of those by H$_2$ torques. 

In molecular clouds where torques due to H$_2$ formation are inefficient due to the lack of atomic hydrogen, grains can rotate suprathermally by the torques discussed in this section, or by isotropic radiative torques. This agrees with observations that indicate that in molecular clouds the temperatures of dust and gas differ (see Lazarian, Goodman \& Myers 1997).

\subsection{Comparison to Anisotropic Radiative Torques}
\subsubsection{Anisotropic Radiative Torques}

\begin{figure}
\includegraphics[width=0.5\textwidth]{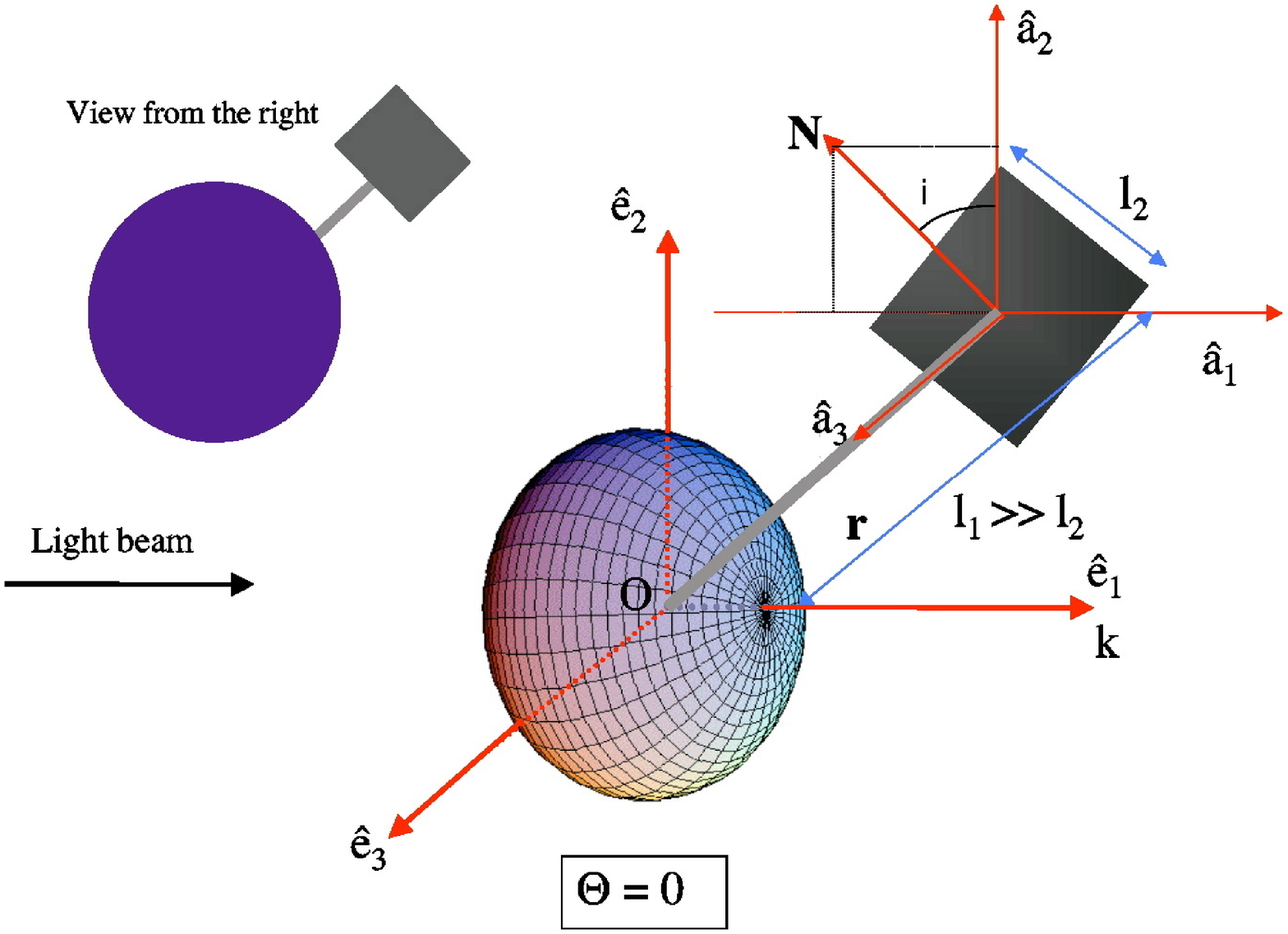}
\includegraphics[width=0.5\textwidth]{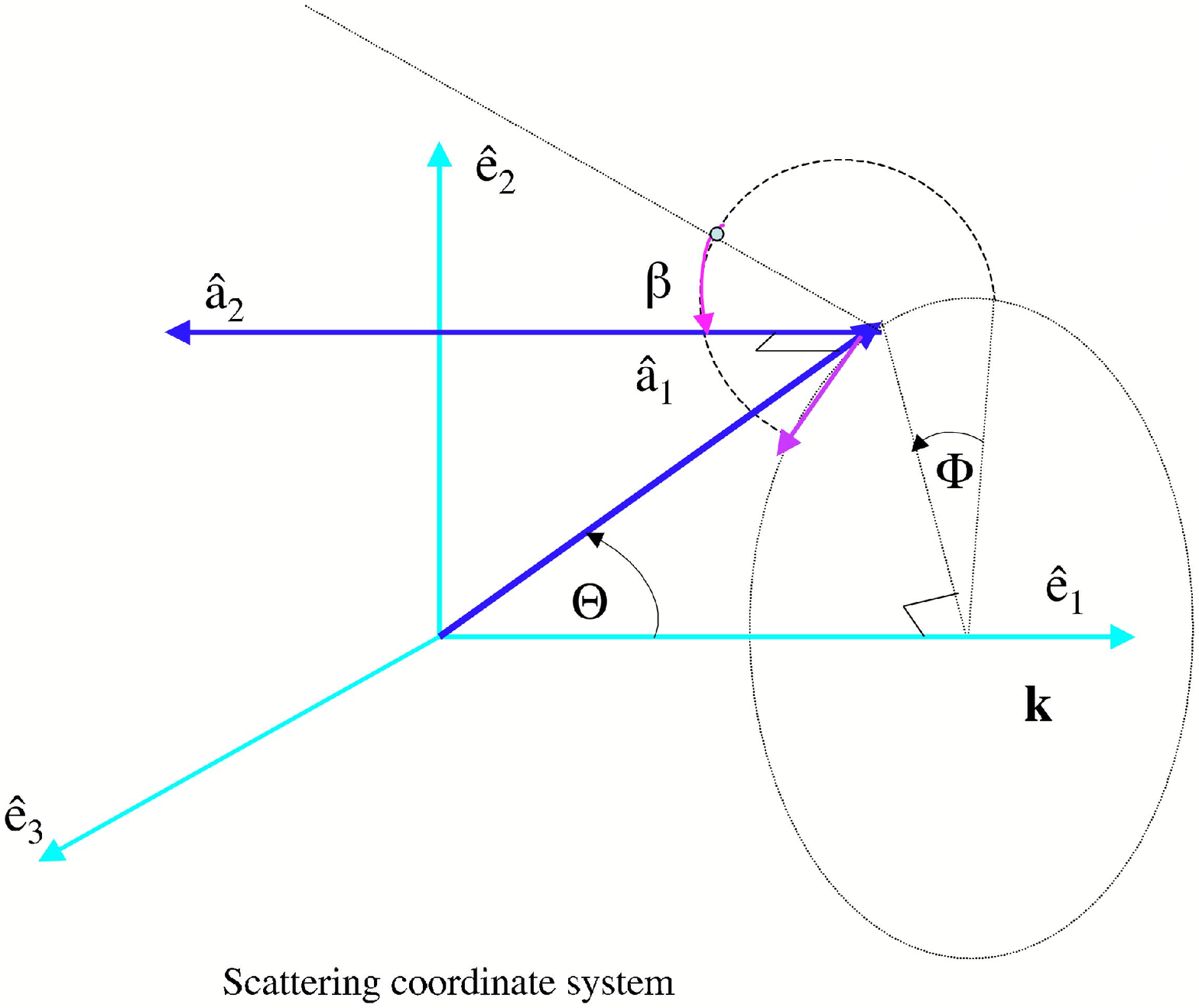}
\caption{{\it Upper panel:} Geometry for AMO consisting of a perfectly reflecting spheroid and a weightless square mirror of side $l_{2}$ tilted by an angle $i$, connected to the spheroid by a rod with length $l_{2}\gg l_{2}$. {\it Lower panel:} The orientation of a grain, described by three principal axes $\hat{a}_{1},\hat{a}_{2}, \hat{a}_{3}$, in the laboratory coordinate system (scattering reference system)  $\hat{e}_{1},\hat{e}_{2}, \hat{e}_{3}$ is defined by three angles $\Theta, \beta, \Phi$. The direction of incident photon beam ${\bf k}$ is along $\hat{e}_{1}$.} 
\label{amo}
\end{figure}

An analytical model (AMO) of RATs arising from the perfect reflection of the radiation beam onto the surface of a helical grain was proposed in LH07a. This grain model consists of a perfect reflecting mirror, connected to an ellipsoid (see the upper panel in Figure \ref{amo}). LH07a proved that the AMO represents very well the basic properties of RATs for irregular grains obtained from DDSCAT.  
\begin{figure}
\includegraphics[width=0.5\textwidth]{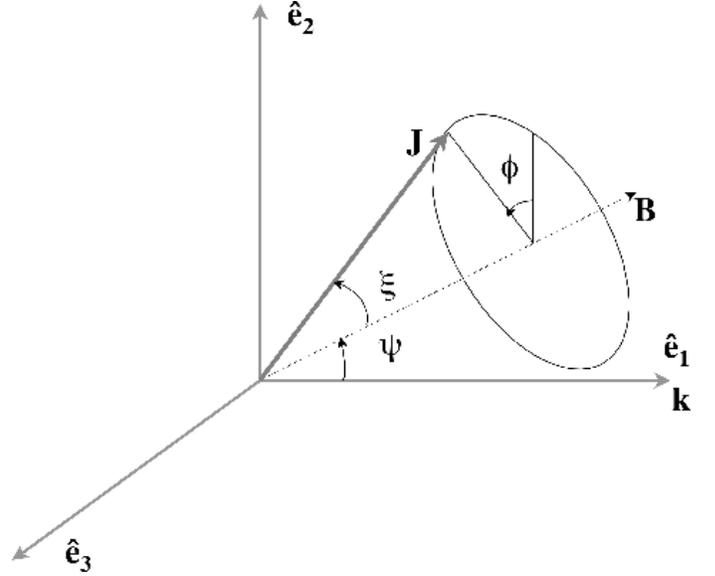}
\caption{Alignment coordinate system: $\psi$ 
  is the angle between the magnetic field {\bf B} and the radiation direction {\bf k}, $\xi$  is the angle between the angular momentum vector ${\bf J}$ and {\bf B}, and $\phi $ is the Larmor precession angle of the magnetic moment parallel to $\bJ$ around ${\bf B}$.} 
\label{sys}
\end{figure}

Throughout the paper, we adopt the functional forms of RATs from the AMO, which are given in Appendix C. To study grain alignment, the amplitude of torque components for the AMO, described by $q^{max}$, can be rescaled to the values for irregular grain obtained using DDSCAT. To account for the magnitude of RATs as functions of grain size and wavelength of radiation, we make use of the approximate scaling obtained in LH07a, which is given by
\bea
Q_{\Gamma}\approx 0.4\left(\frac{{\lambda}}{a}\right)^{\eta},\label{eq9c}
\ena
where $\eta=0$ for $\lambda \sim a$  and $\eta=-3$ for $\lambda \gg a$ (see Appendix C for more detail). 

The magnitude of RATs in dimensionless units is
\bea
J_{max}^{'RAT}&=&{\Gamma'_{RAT}}=\frac{\gamma\bar{\lambda} u_{rad}a^{2} t_{gas}}{2 J_{th}}\overline{Q_{\Gamma}},\\
&\approx&2\times10^{4}\left(\frac{\gamma}{0.1}\right)\left(\frac{\bar{\lambda}}{1.2~\mu m}\right)\left(\frac{u_{rad}}{u_{ISRF}}\right)\hat{\rho}^{1/2}a_{-5}^{1/2}\nonumber\\
&&\times \left(\frac{1}{\hat{n}_{H}\hat{T}_{gas}}\right)\overline{Q_{\Gamma}},\label{rat}
\ena
where $\gamma$ is the anisotropy of radiation field, $\bar{\lambda}$ and $\overline{Q}_{\Gamma}$ are the wavelength and RAT efficiency averaged over the spectrum of radiation field.

Using equation (\ref{eq9c}) for $\overline{Q}_{\Gamma}$ by replacing $\lambda=\bar{\lambda}$, we get
\bea
J_{max}^{'RAT}&=&8\times10^{3}\left(\frac{\gamma}{0.1}\right)\left(\frac{\bar{\lambda}}{1.2~\mu m}\right)^{\eta+1}\left(\frac{u_{rad}}{u_{ISRF}}\right)\hat{\rho}^{1/2}a_{-5}^{1/2}\nonumber\\
&&\times \left(\frac{1}{\hat{n}_{H}\hat{T}_{gas}}\right)\left(\frac{12}{a_{-5}}\right)^{\eta}.\label{rat1}
\ena

\subsection{H$_{2}$ formation torques versus anisotropic RATs}
To study the relative role of pinwheel torques\footnote{In the rest of the paper, we consider H$_{2}$ torques as a representative of pinwheel torques.} to RATs, let us define $\Gamma_{H2}/\Gamma_{RAT}$ to be the ratio of magnitude of  H$_{2}$ torques to RATs. Using equations (\ref{eq12}) and (\ref{rat1}), we obtain
\bea
\frac{J_{max}^{'H_{2}}}{J_{max}^{'RAT}}&=&\frac{\Gamma'_{H_{2}}}{\Gamma'_{RAT}}\approx0.4\left(\frac{0.1}{\gamma}\right)\left(\frac{a_{-5}}{12}\right)^{\eta}\left(\frac{1.2 \mu m}{\bar{\lambda}}\right)^{\eta+1}\frac{u_{ISRF}}{u_{rad}}\nonumber\\
&&\times~ \hat{n}_{H}\hat{T}_{gas}^{1/2} \hat{s}^{5/2}(1-y)\hat{\gamma}_{H}\left(\frac{10^{11}\mbox{ cm}^{-2}}{\alpha}\right)^{1/2}\hat{E}_{kin}^{1/2}.\nonumber\\
\label{eq9e}
\ena

The H$_{2}$ torque-RATs ratio depends on both the grain size $a$ and the active site density $\alpha$, which is a rather uncertain parameter. Therefore, we consider a range of $\alpha$ with the lower limit $\alpha_{min}\sim 1/4\pi a^{2}$, corresponding to one active site on the entire grain surface, and the upper limit $\alpha=10^{15}$ cm$^{-2}$, which is a reasonable value for high density of active sites. For $a_{-5}=1$ grain, equation (\ref{eq9e}) gives ${J_{max}^{'H_{2}}}/{J_{max}^{'RAT}}\sim 2\times 10^{2}$ and for $\alpha=10^{12}$cm$^{-2}$ using standard parameters of the ISM.

\begin{figure}
\includegraphics[width=0.5\textwidth]{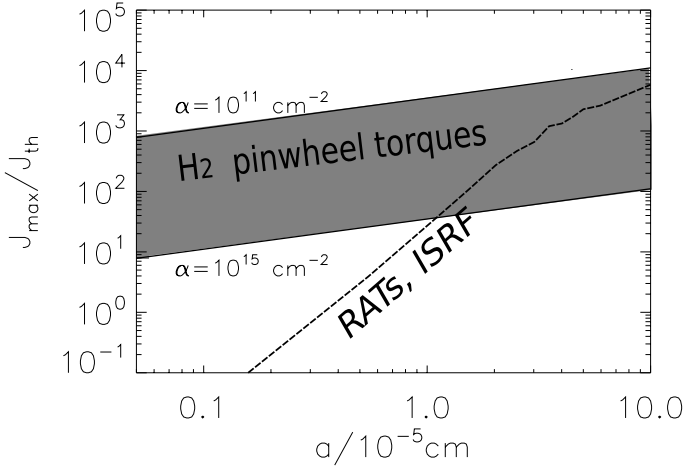}
\caption{The maximum value of angular momentum relative to the thermal value that H$_{2}$ pinwheel torques and RATs can achieve as a function of grain size for the ISM. The shaded area represents the pinwheel torques due to the H$_{2}$ formation (eq. \ref{eq12}), and the solid line shows represents RATs for an irregular grain (shape 1) and the ISRF. For H$_{2}$ pinwheel torques, the upper boundary corresponds to one site per a grain of $a=10^{-6}$cm, and the lower boundary corresponds to an reasonable value for high density of active sites. For RATs, radiation direction is assumed to be parallel to the magnetic field.} 
\label{f4}
\end{figure}

Let us calculate the ratio of maximal angular momentum to its thermal value for the H$_{2}$ formation (eq. \ref{eq12}), and for the anisotropic RATs $Q_{\Gamma}$ calculated using DDSCAT for a more realistic grain (shape 1). $\overline{Q}_{\Gamma}$ is obtained by averaging $Q_{\Gamma}$ over the spectrum of ISRF as in Mathis, Mezger \& Panagia (1983). Figure \ref{f4} shows $J_{max}/J_{th}$ for these two torques. Here we assume that radiation is parallel to the magnetic field and disregard the decrease of RATs due to thermal wobbling of the grain. The shaded area shows $J_{max}/J_{th}$ for H$_2$ torques where the upper and lower boundaries correspond to the lower and upper limits of density of active sites $\alpha$. 

It can be seen that $J_{max}/J_{th}$ for RATs increases with grain size faster than that due to H$_{2}$ pinwheel torques, and becomes dominant for grains larger than $5\times 10^{-5}$cm. In contrast, for grains smaller than $7\times 10^{-6}$cm, H$_{2}$ pinwheel torques are dominant over RATs. \footnote{For $a_{-5}=1$ grain, Figure \ref{f4} shows that ${J_{max}^{'H_{2}}}/{J_{max}^{'RAT}}\sim 60$ for $\alpha=10^{12}$cm$^{-2}$. This value is much smaller than that given by equation (\ref{eq9e}). Such difference stems from the fact that $\overline{Q}_{\Gamma}$ for the ISRF is larger than that given by equation (\ref{eq9c}) used to derive equation (\ref{eq9e}).}

In addition, such small grains can still rotate suprathermally with $J_{max}\gg J_{th}$ by H$_{2}$ torques. In principle, the suprathermal rotation of small grains can produce a detectable degree of alignment as a result of paramagnetic dissipation. This contradicts to observations that grains smaller than $5\times 10^{-6}$cm can not be aligned (Kim \& Martin 1995). This contradiction can be resolved if grains undergo thermal flipping and get thermally trapped, as we discuss below.

We note that the value of RATs for a given grain size, depends on the angle $\psi$ between the magnetic field ${\bf B}$ and the radiation
direction ${\bf k}$. Our study in HL08b shows that the value of the spin-up component of RATs at the angle $\psi$ approaching $\pi/2$ may
be $10^{-2}$ of the value suggested by equation~(\ref{eq9e}) making RATs weaker than the pinwheel torques for some angles $\psi$.

\section{Thermal Flipping and Trapping: Revisiting Lazarian \& Draine 1999}
\subsection{Barnett, Nuclear and Superparamagnetic Internal Relaxation}

Barnett (1915) pointed out that a rotating paramagnetic body gets magnetized with the magnetic moment parallel to the angular velocity\footnote{This is an inverse of the Einstein-de Haas effect, that was used
to measure the spin of the electron. To grain alignment theory the Barnett effect was introduced by Dolginov \& Mytrophanov (1976), who noticed that the effect should induce the magnetic moment of grains.}. This effect can be easily understood using a classical model. 

Consider a paramagnetic grain which is rotating with the angular velocity ${\bf \Omega}$. As the grain rotates with angular velocity ${\bf \Omega}$, the torque acting on the electron spin is  $\frac{d}{dt} {\bf J}={\bf \Omega}\times {\bf J}$. The equivalent torque can be induced by a magnetic field ${\bf H}_{B}$ acting on the magnetic moment $\mu$ associated with the spin, i.e. $1/c {\bf \mu}\times {\bf H}_{B}$, which provides the Barnett-equivalent field
 $H_{eqv}=\frac{J c}{\mu} \Omega$, which can also be presented as 
\bea
{\bf H}_{eqv}=\frac{\Omega}{\gamma_{g}}.\label{bar}
\ena
where  $\gamma_{g}=\mu/(Jc)=eJ/(m_{e}Jc)=e/(m_{e}c)$, 
with $e$ electron charge and $m_{e}$ electron mass is the magneto-mechanical ratio of an electron. 

\begin{figure}
\includegraphics[width=0.5\textwidth]{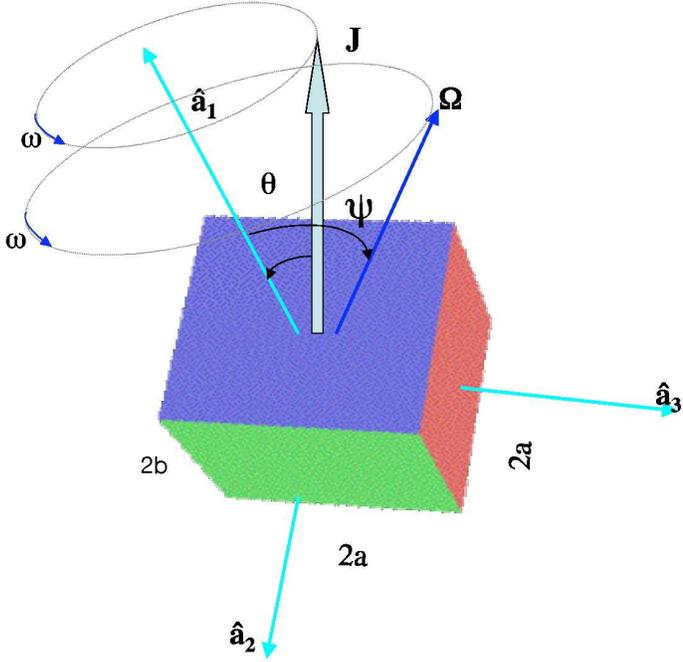}
\caption{Grain model used for calculations of internal relaxation.}
\label{brick}
\end{figure}

Purcell (1979) showed that as the grain wobbles, the changes of magnetization in the grain axes cause the internal relaxation, which he termed "Barnett relaxation". LD99b revisited the problem by
taking into account both spin-lattice and spin-spin relaxation (see Morish 1980). The resulting rate of change for the angle $\theta$ between the grain axis of maximum moment of inertia ${\bf a_1}$ and ${\bf J}$ (see
Figure \ref{brick}) is
\bea
\frac{d\theta}{dt}=-\frac{VK(h-1)J^{2}}{I_{\|}I_{\perp}^{2}\gamma^{2}_{g}}\ms\theta\mc\theta f(\theta, \omega)
\ena
where $V$ is the volume of the grain, $h=I_{\|}/I_{\perp}$, $K=\chi''_{e}/\omega_{1}=\chi(0)\tau_{el}/[1+(\omega_{1}\tau_{el})^{2}]^{2}$ being the imaginary part of the magnetic susceptibility with $\omega_{1}=J\mc\theta/I_{\|}$, $\chi(0)=4.2\times10^{-2}f_{p}\hat{T}_{d}^{-1}$ with $f_{p}$ being the fraction of paramagnetism material in the grain, and $\tau_{el}$ is the relaxation time of electronic spins (Draine 1996; LD99a). $f$ is the factor of order of unity, and when $f\equiv 1$ the result coincides with that of Purcell (1979). Ignoring the
factor $f$ in the above equation, one gets the Purcell (1979) expression for the Barnett relaxation rate for a brick as in Figure \ref{brick} with $I_{\|}=16/3\rho a^{4}b, h=2a^{2}/(a^{2}+b^{2})$:
\bea
t_{Bar}&=&\frac{\gamma_{g}^{2}I_{\|}^{3}}{VKh^{2}(h-1)J^{2}}\nonumber\\
&\approx&2\times 10^{8} \hat{\rho}^{2}\hat{a}_{-5}^{7}\hat{s}[0.5+0.125\hat{s}^{2}]^{2}\left(\frac{J_{d}}{J}\right)^{2}\nonumber\\
&\times& \left[1+\left(\frac{\omega_{1}\tau_{el}}{2}\right)^{2}\right]^{2}\mbox{s},\label{tbar}
\ena
where $\tau_{el}\sim \tau_{2}\sim 2.9\times 10^{-12}f_{p}^{-1}$s with assumption of $f_{p}=0.1$ is the spin-spin coupling time, $\hat{\rho}=\rho/(3$ gcm$^{-3}$) and $\hat{a}_{-5}=a/10^{-5}$ cm; $J_{d}=\sqrt{I_{\|}k_{B}T_{d}/(h-1)}$ is the dust thermal angular momentum \footnote{The relaxation of electron spins results from the spin-latice and spin-spin relaxation, with time scales $\tau_{1}\gg\tau_{2}$, so here we adopted $\tau_{el}\sim \tau_{2}$ (Draine 1996)}.

Although Purcell (1979) considered grains with both electron and nuclear spins, his study missed the effect of internal relaxation related to nuclear spins.
LD99b found that for astrophysical grains of realistic composition nuclear spins induce a new type of relaxation, which was termed by LD99b "nuclear relaxation". This relaxation can be understood
in a simple-minded approach in terms of much stronger {\it equivalent} magnetic field given by equation~(\ref{bar}). Indeed, this field is proportional to the mass of the species involved. As the paramagnetic relaxation is proportional to $H^2_{eqv}\chi'' \varpropto\gamma_{g}^{2}\chi(0)\tau\varpropto m^{2}(1/m^{2})\tau\sim \tau$, and as $\tau_{n}\gg\tau_{el}$, one can understand the nature of the dominance of the nuclear relaxation, which has a characteristic time:
\bea
t_{nucl}&\approx&6\times 10^{2}\hat{\rho}^{2}\hat{a}_{-5}^{7}\left(\frac{J_{d}}{J}\right)^{2}\hat{s}[0.5+0.125\hat{s}^{2}]^{2}\nonumber\\
&\times&\left[1+\frac{(\omega_{1}\tau_{n})^{2}}{2}\right]^{2}\mbox{ s},
\ena
where $\tau_{n}^{-1}=\tau_{nn}^{-1}+\tau_{ne}^{-1}$ is the relaxation rate induced by the nuclei-nuclei and electron-nuclei spin interactions (see LD99b). 

Starting with the classical study by Jones \& Spitzer (1968) superparamagnetic inclusions are frequently considered as a constituent part of astrophysical dust grains (see Mathis 1986; Martin 1994,
Goodman et al. 1995; Draine \& Lazarian 1999; Roberge \& Lazarian 1999). However, the effect of superparamagnetic inclusions on the internal relaxation was considered only recently, namely in
Lazarian \& Hoang (2008, LH08). There it was shown that superparamagnetic inclusions enhance the efficiency of the Barnett relaxation and also extend the range of frequencies for which the nuclear 
relaxation is efficient. In particular, the superparamagnetically enhanced rate of the Barnett relaxation is 
\bea
t_{Bar,sup}&\approx&10^{8}\hat{\rho}^{2}\hat{a}_{-5}^{7}\frac{1}{N_{cl}}\mbox{exp}\frac{-0.011N_{cl}}{T_{d}}\left(\frac{J_{d}}{J}\right)^{2}\nonumber\\
&&\times\hat{s}[0.5+0.125\hat{s}^{2}]^{2}\left[1+\left(\frac{\omega_{1}\tau_{sup}}{2}\right)^{2}\right]^{2}\mbox{ s},\label{tau}
\ena
where $\tau_{sup}^{-1}=10^{9}\mbox{exp}(-N_{cl}T_{act}/T_{d})$ with $N_{cl}$ being the number of iron atom per cluster and $T_{act}=0.011$K is the rate of remagnetization due to superparamagnetic inclusions, and the fraction $f_{sup}=0.01$ of atoms being magnetic was adopted (see LH08).

In addition, Purcell (1979) considered the internal relaxation arising from grains being inelastic. The improved rates of this relaxation were obtained in Lazarian \& Efroimsky (1999). However, for the 
typical grains in the diffuse interstellar medium, this rates are subdominant. The inelastic relaxation gets important for large grains, which we consider in other papers (e.g. in HL08b).

\begin{figure}
\includegraphics[width=0.5\textwidth]{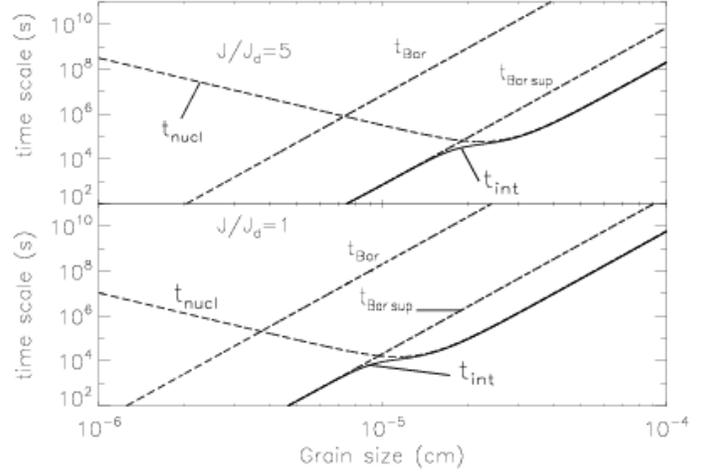}
\caption{Timescales for different internal relaxation processes as a function of the grain size for $J=5J_{d}$ and $J_{d}$. The number of iron atom per cluster $N_{cl}=2\times 10^{3}$ is assumed.} 
\label{f4a}
\end{figure}

All in all, the total internal relaxation rate $t^{-1}_{int}(J)$ is the sum of all relaxation rates
\bea
t^{-1}_{int}\approx t^{-1}_{Bar}+t^{-1}_{Bar,sup}+t^{-1}_{nucl}+...,\label{tint}
\ena
where ... are placed instead of other relaxation rates, e.g. inelastic relaxation rate. 
Figure \ref{f4a} shows the timescales for various internal relaxation processes as a function of the grain size for $J=5J_{d}$ and $J_{d}$. We assume $N_{cl}=10^{4}$ for superparamagnetic inclusions. 

\subsection{Revised model of thermal flipping}\label{flip}

Internal relaxation induces the thermal wobbling (Lazarian 1994; Lazarian \& Roberge 1997), which according to LD99a can result in interesting effects like thermal flipping and thermal trapping. An
alternative, but consistent with LD99a treatment of the thermal flipping was introduced in a preprint by Roberge \& Ford (2000; hereafter RF00). Both treatments use the diffusion coefficients derived in
Lazarian \& Roberge (1997). However, Weingartner (2008) questioned both the derivation of the diffusion coefficients and the phenomena of thermal flipping and thermal trapping. 

Here we introduce a model that generalizes the one in LD99a, namely, it takes into account the fact that the processes that induce pinwheel torques also induce stochastic torques. Such torques should be taken into account in the treatment of suprathermally rotating grains. In addition, astrophysical grains are embedded in gas, that also induces stochastic bombardment. Formally, the diffusion coefficients arising from the internal relaxation and external stochastic torques are different. For instance, internal relaxation cannot change the angular momentum, but external torques can do so.
However, within the formalism that we are considering, the external stochastic torques are subdominant for all $\theta$ angles apart from $\theta\rightarrow \pi/2$. Therefore, for the sake of simplicity, we disregard the small change of the angular momentum amplitude that the external stochastic torques cause. Instead, we concentrate on their influence on the variation of the angle $\theta$.\footnote{Ideology of this approach is similar to that employed by Spitzer \& McGlynn (1979) and Lazarian \& Draine (1997) while treating grain crossovers.}

Combining dimensionless diffusion coefficients for internal relaxation, and impulses due to $H_{2}$ formation and gas bombardment, we get
\bea
B'_{total}(\theta)=B'(\theta)+\frac{t_{gas}}{t_{int}}\left(B'_{H_{2}}+B'_{coll}\right),\label{eq24}
\ena
where $B'(\theta)$ and $B'_{H_{2}}$ are given by equations (\ref{bbar}) and (\ref{bh2}), and $B'_{coll}$ is obtained using equations (\ref{bh2}) combined with (\ref{col1})-(\ref{col3}).
The relative role of internal relaxation and H$_{2}$ formation and gas bombardment  are described by the ratio $t_{gas}/t_{int}$. 

Our treatment of grain flipping is similar to the treatment in RF00\footnote{The treatment in LD99a is based on the Zeldovich approximation to the first order transitions. It can also be generalized to the
presence of the external torques, but the generalization is somewhat involved.}, where the approach to solve analytically the stochastic differential equations was adopted.  That
approach was described in the textbook of Gardiner (1983, \S 5.2.7). For the sake of completeness, we discuss this approach in Appendix B.
Gardiner (1983) described the problem of the escape of a particle initially confined within the range $[a,b]$ with $x=a$ being a reflecting boundary and $x=b$ being an absorbing boundary. 
In our flipping problem, the escape corresponds to the angle $\theta$ passing $\pi/2$ value. When $\theta$ gets larger than $\pi/2$, following RF00, we claim the flipping event. 

In the following, we adopt this approach but study the effect of thermal flipping in the presence of impulses due to H$_{2}$ formation and gas bombardment. As we discussed earlier, the internal relaxation is fast, so it is reasonable to assume that angular momentum is constant in the presence of external torques. We calculate the mean flipping timescale for all internal relaxation processes. For the superparamagnetic case, the number of iron per cluster $N_{cl}=2\times 10^{3}$ is adopted. Because the mean flipping time depends on the initial angle $\theta$ (see eq. \ref{eqa7}), here we show the mean flipping time for $\theta=30^{\circ}$. 
\begin{table}
\caption{Physical parameters used for calculations in the paper\label{tab1}}
\begin{tabular}{lll} \hline\hline\\
\multicolumn{1}{c}{\it Parameters} & \multicolumn{1}{c}{\it Diffuse~ISM}& {\it MCs}\\[1mm]
\hline\\
{\rm n$_{H}$(cm$^{-3}$)}& {\rm $30$} &{\rm $300$}\\[1mm]
{\rm T$_{gas}$(K)}& {\rm 100}& {\rm 20}\\[1mm]
{\rm T$_{d}$(K)}& {\rm 15}& {\rm 15}\\[1mm]
{\rm y=2n(H$_{2}$)/n$_{\mbox{H}}$}&{$1$}&{$10^{-3}$}\\[1mm]
{\rm $\gamma$}&{\rm 0.1}&{\rm 0.1}\\[1mm]
{\rm $\overline{\lambda}$}&{\rm 1.2}&{\rm 1.2}\\[1mm]
{\rm ${u}_{rad}$}&{\rm $8.64\times10^{-13}$}&{varied}\\[1mm]
{\rm $\gamma_{H}$}& {\rm 0.1}& {\rm 0.1}\\[1mm]
{\rm t$_{gas}$(s)}& {\rm $2.3\times 10^{12}(\frac{\hat{\rho}}{\hat{n}\hat{T}_{gas}^{1/2}}) a_{-5}$} &{\rm the~ same}\\[1mm]\\[1mm]
\hline\hline\\
\end{tabular}
\tablecomments{Here $\hat{T}_{gas}=T_{gas}/100~ \mbox{K},~\hat{n}=n/30~ \mbox{g cm}^{-3}$, and
$a_{-5}=a_{eff}/10^{-5}~ \mbox{cm} $.~ $\hat{\rho}=\rho/3~ \mbox{g}~ \mbox{cm}^{-3}$ where $\rho=3~ \mbox{g cm}^{-3}$ is the mass density of the grain.}
\end{table}
\begin{figure}
\includegraphics[width=0.5\textwidth]{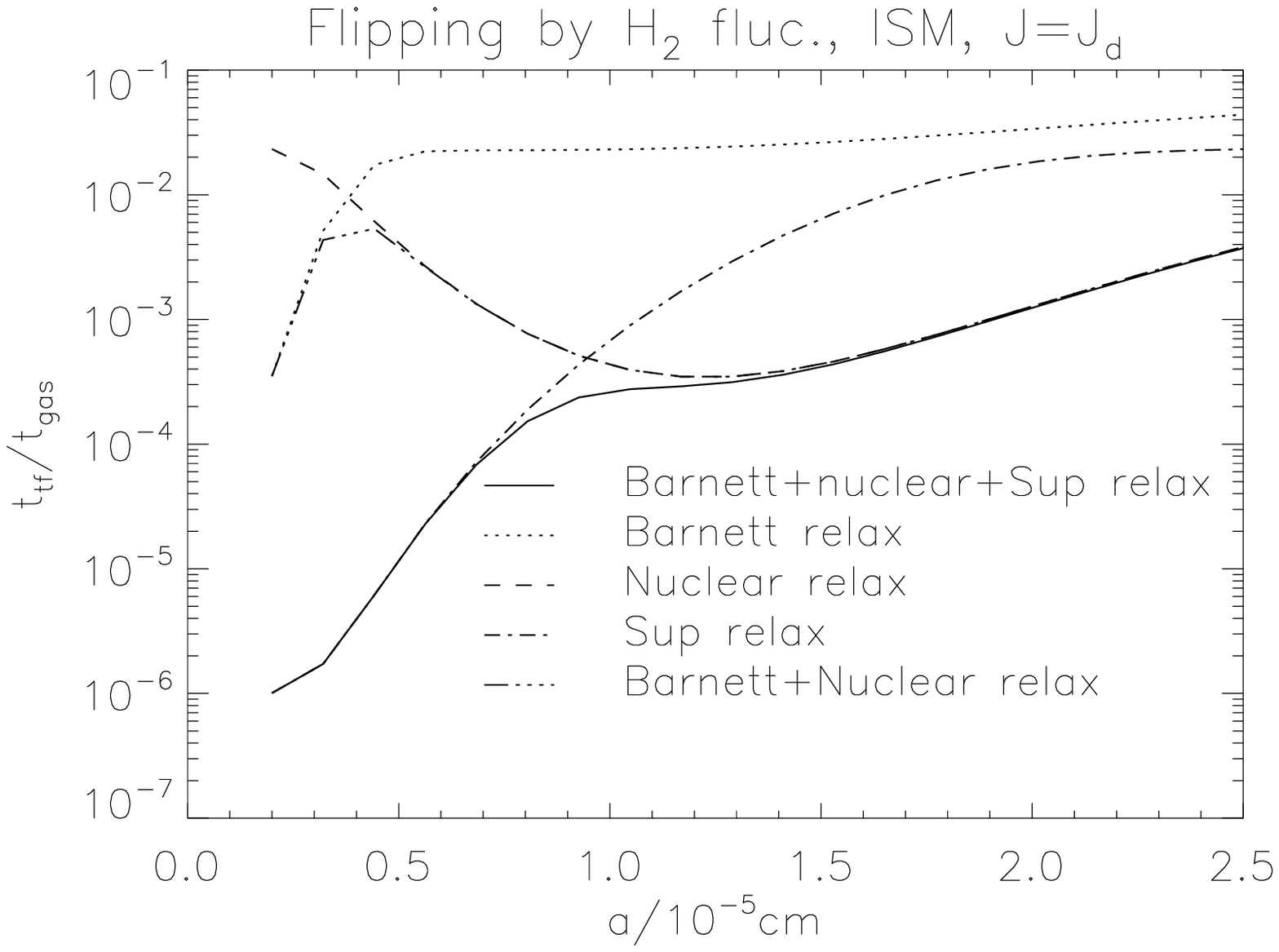}
\includegraphics[width=0.5\textwidth]{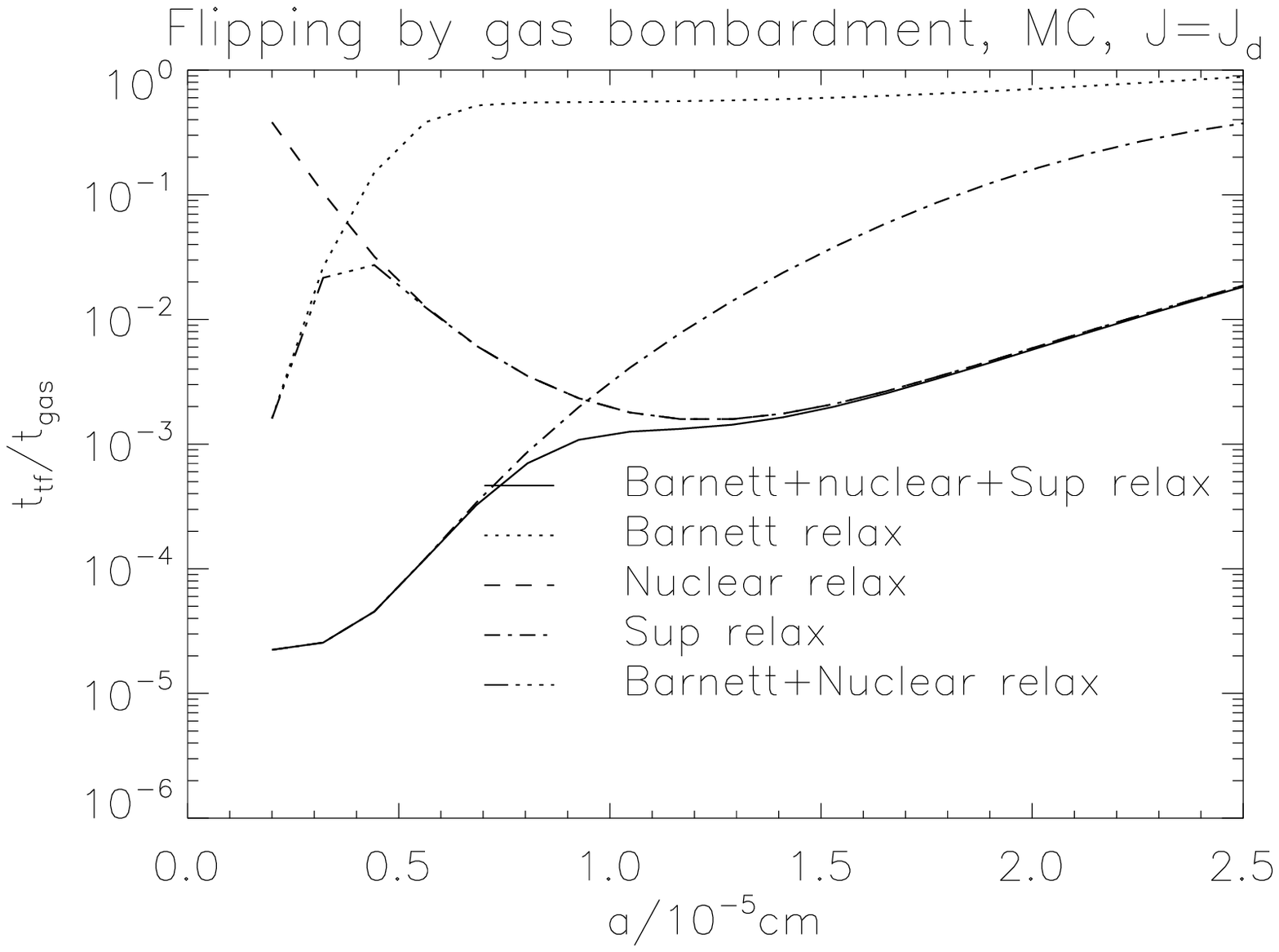}
\caption{Mean flipping time relative to the gas damping time of grain induced by internal relaxation in the presence of impulses associated with H$_{2}$ formation for the ISM ({\it upper}) and gas bombardment for molecular cloud ({\it lower)}. The angular momentum is assumed to be constant and $J=J_{d}$. Physical parameters used in calculations are given in Table 1.}
\label{f5}
\end{figure}

When we disregard the external stochastic torques, we find, in agreement with Weingartner (2008), that the mean flipping time is infinite, i.e., the grain can not flip by internal relaxation only. However, for any realistic rates of external, e.g. stochastic torques, the flipping is present. 

Figure \ref{f5} shows the ratio of mean flipping time to the gas damping time, $t_{tf}/t_{gas}$, as a function of the grain size $a$ for the diffuse ISM (upper panel) and a molecular cloud (MC; lower panel). For the ISM, the stochastic torques due to $H_{2}$ formation are dominant because the number density of atomic hydrogen $n(H)$ is comparable to the total number density $n_{H}$.\footnote{We disregard the effect of gas bombardment in the ISM because of the dominance of H$_{2}$ impulses, i.e., $B_{H_{2}}^{'}=E_{kin}/4k_{B}T_{gas}\sim 0.2 \times 1.6 10^{-12}/(4\times 1.38\times 10^{-14})\sim 6 > B_{'coll}\approx 1$.} But in the MC, most hydrogen is in molecular state, i.e., $n_{H}\sim n(H_{2})$, so we can disregard impulses due to $H_{2}$ formation and account only for gas bombardment.

For smaller grains, the internal relaxation is much stronger than the external impulses, $t_{tf}/t_{gas}$ changes rapidly with the grain size $a$, and the curves for $t_{tf}/t_{gas}$ for different relaxation processes follow the steep slopes as seen with $t_{int}$. For larger grains, the internal relaxation gets weaker, and $t_{tf}/t_{gas}$ exhibits slower increases with $a$.

\subsection{Critical size of flipping}\label{trap}
Let us consider first whether the grain undergoing spin-down due to pinwheel torques experiences a regular crossover in the fashion described in Spitzer \& McGlynn (1979) and Lazarian \& Draine (1997) or
it undergoes a flip-over. The regular crossover time according to Lazarian \& Draine (1997) is
\bea
t_{cros}=\frac{2J_{\perp}}{d{J}_{\|}/dt}=\frac{2J_{\perp}}{\Gamma_{pinw}},\label{eq30}
\ena
where $\Gamma_{pinw}$ is the magnitude of pinwheel torques (see $\S 2$). Here we consider the pinwheel torques due to H$_{2}$ formations as a representative for various kinds of pinwheel torques discussed  in $\S 2$.  We also assume that $J_{\bot}=J_{d}$ as in Lazarian \& Draine (1997). Then equation (\ref{eq30}) can be rewritten
\bea
\frac{t_{cros}}{t_{gas}}=\frac{2}{J'_{max}}\left(\frac{J_{d}}{J_{th}}\right)=\frac{2}{J'_{max}}\left(\frac{T_{d}}{(h-1)T_{gas}}\right)^{1/2},\label{eq30b}
\ena
where $J'_{max}=J_{max}^{'H_{2}}$. In the following, we consider only the pinwheel torques due to H$_{2}$ formation. It is trivial to generalize the treatment for other pinwheel torques. The crossover time given by equation (\ref{eq30b}) will be compared with the mean flipping time $t_{tf}$ to get the critical size of flipping.

Using the pinwheel torque due to H$_2$ formation (eq. \ref{eq12}) with the use of equation (\ref{eq30b}), we can calculate $t_{cros}$ for various grain sizes $a$ and active site density $\alpha$. For a given $\alpha$, the critical size of flipping, $a_{cri}$ is then derived by setting $t_{cros}=t_{tf}$, where $t_{tf}$ are calculated in \S~3.2.
\begin{figure}
\includegraphics[width=0.48\textwidth]{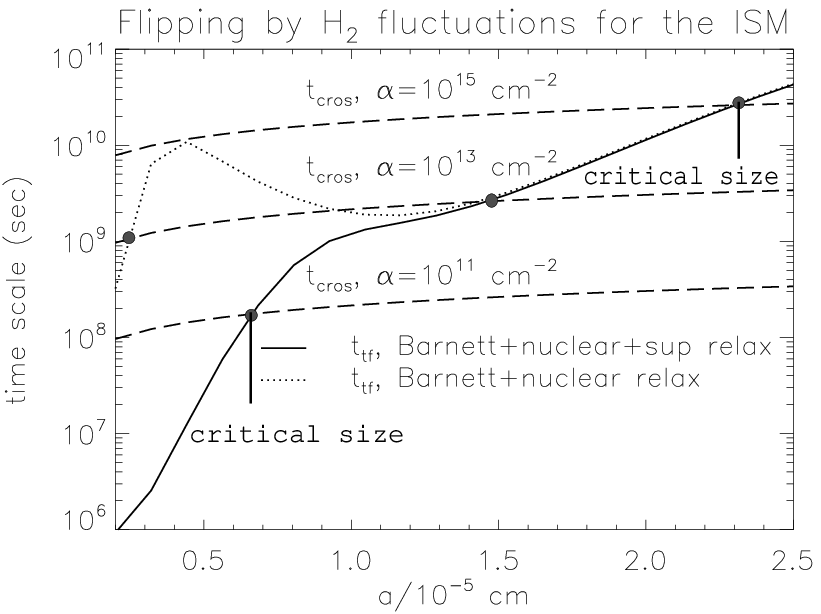}
\includegraphics[width=0.52\textwidth]{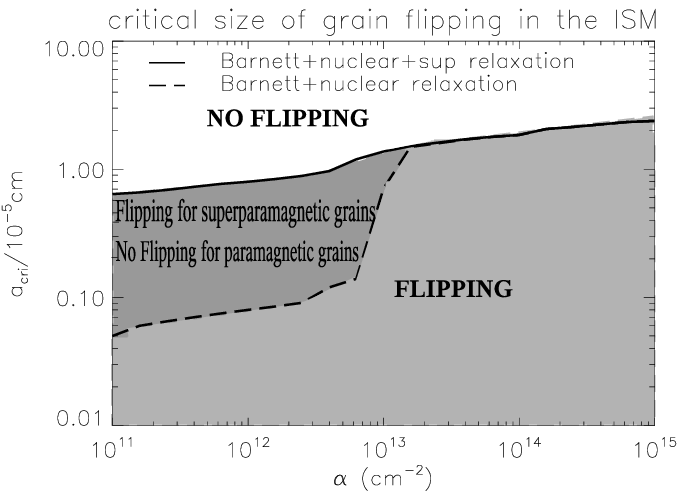}
\caption{{\it Upper:} Mean flipping time compared to crossover time as a function of grain size for three values of active site density, $\alpha$. The grain size at the cross of two timescales defines a critical size $a_{cri}$ of grain flipping. {\it Lower:} Obtained critical size for different  $\alpha$ for Barnett and nuclear relaxation (dashed line) and total internal relaxation (solid line). Regions where grains can and can not flip are denoted with flipping and no flipping.}
\label{f7}
\end{figure}

The upper panel in Figure \ref{f7} shows the mean flipping time $t_{tf}$ in the presence of stochastic H$_{2}$ torques, and crossover time $t_{cros}$ for three values of active site density $\alpha=10^{11}, 10^{13}$ and $10^{15}$ cm$^{-2}$. The solid line shows $t_{tf}$ when all internal relaxation processes (Barnett, nuclear and superparamagnetic inclusions) are accounted for, and the dot line represents that when only  Barnett and nuclear relaxations are taken into account. The filled circles denote the intersects of $t_{tf}$ and $t_{cros}$, i.e. where $t_{tf}=t_{cros}$. The grain size corresponding to this intersect is called critical size of flipping, $a_{cri}$. For $\alpha$ in the range of $10^{11}$ to $10^{15}$ cm$^{-2}$, we can estimate $a_{cri}$ as a function of $\alpha$.


The obtained results are plotted in Figure \ref{f7} when different internal relaxation processes are accounted for and in the presence of stochastic $H_{2}$ torques. 
The region where grains with $a< a_{cri}$ can flip is denoted with flipping, and the region in which grains with $a>a_{cri}$ can not flip is denoted with no flipping.

It can be seen that for $\alpha> 10^{13}$cm$^{-2}$, $a_{cri}$ is similar for both paramagnetic and superparamagnetic grains. But for $\alpha< 10^{13}$cm$^{-2}$, $a_{cri}$ for the former decreases substantially, and reaches very small size $\sim 5\times 10^{-7}$cm for $\alpha=10^{12}$cm$^{-2}$, while $a_{cri}$ is about $6\times 10^{-6}$cm for the later case. Therefore, it exists an area, A, in which the superparamagnetic grains can, but paramagnetic grains can not flip.

\subsection{Thermal Trapping}

In this section we follow LD99a in describing the problem of thermal trapping. This effect happens when the flipping is fast, so that the regular pinwheel torques cannot appreciably accelerate the grain between
the flips. As flips happen, the regular pinwheel torques change the sign and are averaged to zero. We consider the flipping induced by total internal relaxation, in the presence of impulses due to H$_{2}$ formation for the ISM. 

Following LD99a, whether the grain can be spun-up to suprathermal rotation after the crossover or not depends on the acceleration by the systematic torques and the thermal mean flipping time. Let $J_{trans}$ be the angular momentum at which the acceleration time by systematic torques is equal to the mean flipping time. We obtain
\bea
\frac{t_{tf}(J'_{trans})}{t_{gas}}=\frac{J'_{trans}}{(G-1)^{1/2}},\label{eqq9}
\ena
where $J'_{trans}=J_{trans}/J_{th}$, and $G$ is the total systematic torque given by
\bea
G^{2}=\langle (\Gamma'_{H_{2}})^{2}\rangle^{1/2},\label{eqq9a}
\ena
where  $\langle (\Gamma'_{H_{2}})^{2}\rangle^{1/2}$ is the amplitude of H$_{2}$ pinwheel torques in dimensionless unit given by equation (\ref{eq12}).

When the flipping is rapid, i.e., when thermal trapping occurs, the torques in the grain body coordinate system change randomly. As a result, the mean angular momentum is given by LD99a
\bea
J^{2}=J_{th}^{2}+\frac{(G-1)J_{th}^{2}t_{tf,trap}}{(t_{tf,trap}+t_{gas})},
\label{eqq10}
\ena
where $J_{th}$, as we discussed earlier, is the grain angular momentum corresponding to the gas temperature and $t_{tf,trap}$ is the timescale for the thermal trapping when the grain is trapped at a low-J state by frequent thermal flipping.
From equation (\ref{eqq10}) we obtain
\bea
\frac{t_{tf,trap}}{t_{gas}}=\frac{(J/J_{th})^{2}-1}{G-(J/J_{th})^{2}}.\label{eqq11}
\ena

By setting the mean flipping time $t_{tf}$ obtained from calculations for total internal relaxation  in the presence of H$_{2}$ impulses (Fig. \ref{f5}), equal to that given by equation (\ref{eqq11}), we can obtain solutions for $J$. Let us denote the lower value $J_{1}$ and the upper one $J_{2}$. For $J<J_{1}$, grains are not thermally trapped, and they can be spun-up to $J=J_{1}$. However, for $J_{1}<J<J_{2}$, grains are thermally trapped because the thermal mean flipping time is smaller than the acceleration time. When $J\ge J_{2}$, grains are not thermally trapped at all, and able to rotate suprathermally. Therefore, hereafter, we define $J_{2}$ the critical value of thermal trapping. Note that the degree of alignment depends on the value of angular momentum, i.e on the lower solution $J_{1}$, thus, the variation of $J_{1}$ with the pinwheel  torques is also important.

\begin{figure}

\includegraphics[width=0.5\textwidth]{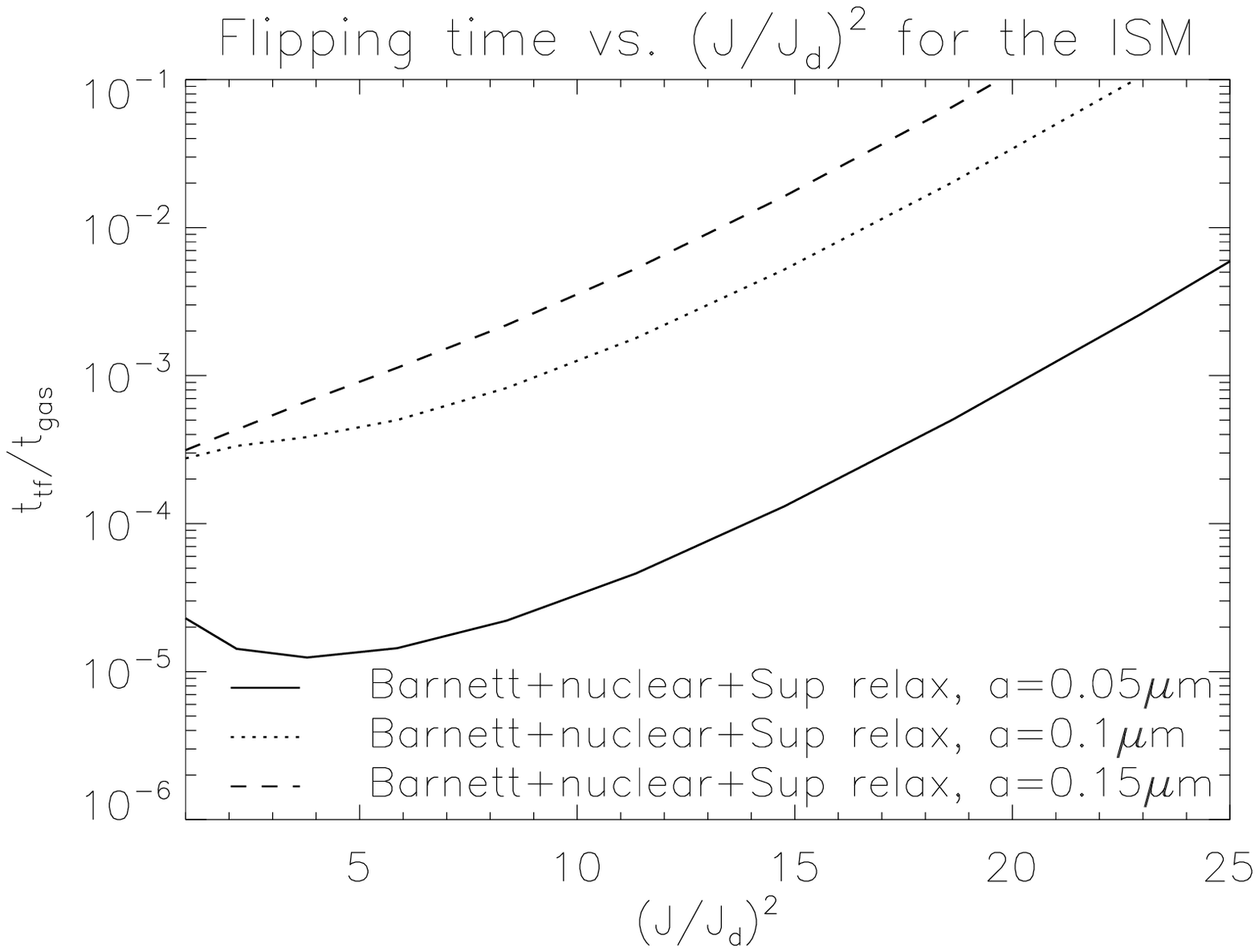}
\includegraphics[width=0.5\textwidth]{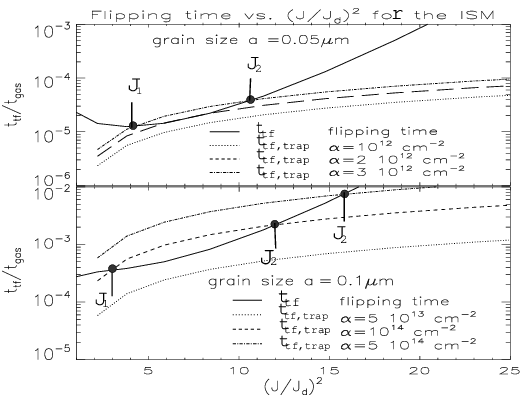}
\caption{{\it Upper panel:} Mean flipping time, $t_{tf}$, induced by total internal relaxation in the presence of impulses due to H$_{2}$ formation, as a function of $(J/J_{d})^{2}$ for the ISM. The result for grain sizes $a=0.05$, $0.1$ and $0.15\mu$m are presented. {\it Lower panel:} $t_{tf}$ similar to the upper panel, and $t_{tf,trap}$ as a function of $(J/J_{d})^{2}$ for different $\alpha$, and for grain sizes $a=0.05$ and $a=0.1 \mu m$. Filled circles depict the angular momenta when the mean flipping time equal the acceleration time. For a given $a$, the decrease of $\alpha$ decreases the acceleration time, and thus decreases the thermal trapping range.}
\label{f8}
\end{figure}

The upper panel of Figure \ref{f8} shows $t_{tf}/t_{gas}$ as a function of $(J/J_{d})^{2}$ for the ISM and for grain size $a=0.05, 0.1$ and $0.15\mu$m. It can be seen that $t_{tf}/t_{gas}$ tends to increase with $J/J_{d}$; larger grains have $t_{tf}/t_{gas}$ increasing faster than small grains with $J/J_{d}$.

In Figure \ref{f8} ({\it lower panel}) we show $t_{tf,trap}$ given by (\ref{eqq11}) for different $\alpha$ and $t_{tf}$ for two grain sizes $a=0.05$ and $0.1 \mu$m. The filled circles show the position of $J_{2}$ and $J_{1}$.  There, it is shown that different grain sizes can be trapped at   $J<J_{2}$ with $J_{2}$ depending on $\alpha$. Small grain $a=0.05 \mu m$ is trapped at $J<J_{2}=\sqrt{11}J_{d}$ even in the presence of very strong pinwheel torques (see the lower panel of Fig. \ref{f8}). 

Following LD99a, we can estimate the trapping time of grains based on the value of $J_{2}$ as
\bea
t_{trap}=t_{gas}\mbox{exp}\left(\frac{J_{2}}{J_{th}}\right)^{2}.\label{ttrap}
\ena

\begin{figure}
\includegraphics[width=0.5\textwidth]{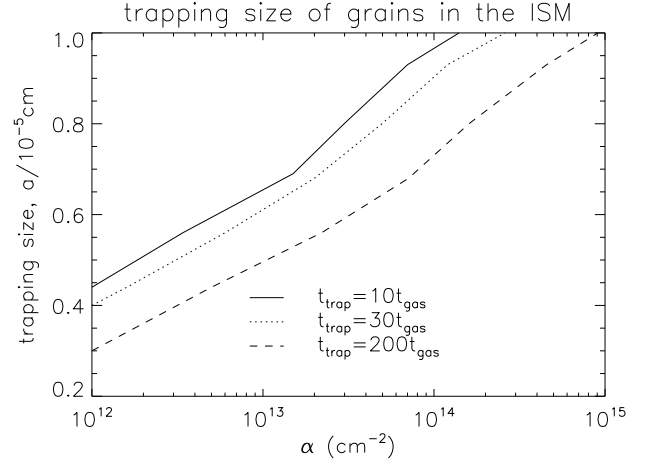}
\caption{Trapping size of the ISM grains as a function of the magnitude of pinwheel torques for trapping time $t_{trap}=10, 30$ and $100 t_{gas}$. Here we assume that the flipping is produced by total internal relaxation in the presence of impulses due to H$_{2}$ formation.}
\label{f8b}
\end{figure}

Because the trapping size depends on both $J_{2}$ and $\alpha$. Using equation (\ref{ttrap}), we can estimate the trapping size for a given trapping time, as a function of $\alpha$, and the results are shown in Figure \ref{f8b}.  The figure shows that grains as small as $3\times 10^{-6}$cm can be thermally trapped for $200 t_{gas}$ by the maximal pinwheel torques with $\alpha=10^{12}$ cm$^{-2}$. For the same trapping time, the trapping size increases with the decrease of pinwheel torque efficiency.

Flipping and thermal trapping were invoked in LD99a to explain why observations testify that the grains with $a<5\times 10^{-6}$~cm are not aligned in the diffuse interstellar gas. Indeed, suprathermally rotating grains are bound to be aligned by the original Purcell (1979) mechanism. The efficiency of the mechanism is somewhat unclear. The original work by Purcell (1979) predicted for short-lived pinwheel torques a rather marginal increase of the alignment compared with the classical predictions in Jones \& Spitzer (1967). This conclusion was corrected by Lazarian \& Draine (1997), who showed that for grains larger than a certain size $a_{cri}$, high degrees of Purcell-type alignment are possible. The size $a_{cri}$ was estimated by Lazarian \& Draine (1997) to be around $2\times 10^{-5}$~cm for the Barnett relaxation. It gets larger than $10^{-4}$~cm if the  nuclear relaxation is accounted for (LD99b). Thus, for small grains, the original conclusions in Purcell (1979) are valid. However, the result in LD99b assumed that the thermal flipping time is proportional to the internal relaxation time. Using the new result for thermal flipping, we found that $a_{cri}$ spams from $10^{-6}$ to $10^{-5}$ cm, which is much smaller than the value obtained in LD99b.

Thus, if sufficiently strong long-lived pinwheel torques dominate the dynamics of small grains, they should exhibit degrees of alignment which contradicts to observations. The problem dissolves if the torques are either weaker than the threshold we discussed above or grains are superparamagnetic or pinwheel torques are short-lived. In first two cases fast flipping of grains induces thermal trapping and grains rotate thermally in spite of the presence of suprathermal torques (see LD99a). If torques are short-lived and grains are smaller than the critical size, the Barnett and nuclear-induced fluctuations randomize grains during crossovers. In this case, the original Spitzer-McGlynn theory should be used and it predicts a marginal alignment.

\section{Alignment by Interstellar RATs plus Pinwheel torques}

In this section we study the alignment of grains by RATs in the diffuse ISM in the presence of H$_{2}$ torques.  We consider the influence of long-lived H$_{2}$ pinwheel torques on the RAT alignment when (i) only low-J attractor points and (ii) both low-J and high-J attractor points exist in the trajectory maps. We study for a range of grain size from $a=0.02$ to $ 0.2\mu$m. Our calculations for thermal flipping showed that the flipping is most efficient for the total internal relaxation (i.e., Barnett, nuclear and superparamagnetic inclusions relaxation) in the presence of  H$_{2}$ formation (see \S 2). Therefore, we adopt this flipping rate in our study below.

\subsection{Equations of motion} 
The motion of a grain can be separated into the torque-free motion of grain axes about angular momentum $\bJ$ and the motion of $\bJ$ about the magnetic field $\bB$ driven by external torques, because the former process is much faster than the later one (see WD03; HL08a). 

The motion of the angular momentum vector $\bJ$ about the magnetic field $\bB$ subjected to a net torque is completely determined by three variables: the angle  $\xi$  between  $\bf$ and 
 {\bf B}, the precession angle $\phi $ of ${\bf J}$ around $\bB$ and the value of the angular momentum $J$ (see Fig. \ref{sys}). The equations of motion of grains by RATs plus H$_{2}$ pinwheel torques, when the paramagnetic dissipation is disregarded, are given by
\begin{eqnarray} 
\frac{d\phi}{dt}&=& \frac{\gamma
  u_{\mbox{rad}}a^{2}\overline{\lambda}}{2J
  \mbox{sin }\xi}\langle G(\xi, \phi, \psi,
J)\rangle-\Omega_{B},\label{eeq15}\\ 
\frac{d\xi}{dt}&=&\frac{\gamma u_{\mbox{rad}}a^{2}\overline{\lambda}}{2J}\langle F(\xi, \phi, \psi,
J)\rangle,\label{eeq16}\\ 
\frac{dJ}{dt}&=&\frac{1}{2}\gamma u_{\mbox{rad}}a^{2}\overline{\lambda} \langle H(\xi, \phi, \psi,
J)\rangle+\Gamma_{H_{2}}.\overline{\mc\theta}-\frac{J}{t_{gas}},\label{eeq17}
\end{eqnarray} 
where $\langle F(\xi,\phi,\psi, J)\rangle, \langle G(\xi,\phi,\psi, J)\rangle$, 
and $\langle H(\xi,\phi,\psi, J)\rangle$ are the RATs components averaged over the torque-free motion and thermal fluctuations (see HL08a); $\Omega_{B}$ is the Larmor precession rate, $a$ is the grain size, $\gamma$ is the degree of anisotropy of radiation, $\overline{\lambda}$ is the mean wavelength of radiation field, $u_{rad}$ is the radiation energy density, and $t_{gas}$ is the gas damping time (see Table 1). Here $\overline{\mc\theta}\propto\int \mc\theta f(\theta, J)\ms\theta d\theta$ with $\theta$ angle between $\ba_{1}$ and $\bJ$ and $f(\theta,J)$ is the  thermal distribution function. $\Gamma_{H_{2}}$ is the component along $\ba_{1}$ axis of H$_{2}$ pinwheel torques.

For the ISM, the Larmor precession rate is larger than the gas damping and the alignment rates; and thus, we can average equations (\ref{eeq16}) -(\ref{eeq17}) over a precession period for $\phi$ from $0$ to $2 \pi$. As a result, equations (\ref{eeq15})-(\ref{eeq17}) can be reduced to a set of equations for $\xi$ and $J$, in which the spinning and aligning torques $\langle F\rangle$ and $\langle H\rangle$ are replaced by $\langle F\rangle_{\phi}$ and $\langle H\rangle_{\phi}$, resulted from averaging corresponding RATs components over the Larmor precession angle $\phi$.\footnote{For the sake of simplicity, hereafter, we denote $\langle F\rangle =\langle F(\xi,\phi,\psi, J)\rangle$, $\langle H\rangle =\langle H(\xi,\phi,\psi, J)\rangle$, and $\langle F\rangle_{\phi} =\langle F(\xi,\phi,\psi, J)\rangle_{\phi}$, $\langle H\rangle_{\phi} =\langle H(\xi,\phi,\psi, J)\rangle_{\phi}$.} In dimensionless units ($J'=J/J_{th}, t'=t/t_{gas}$) and taking into account the grain flipping, we obtain
\bea
\frac{d\xi}{dt'}&=&\frac{M}{J'}\left(\langle F\rangle_{\phi,+}f_{+}+\langle F\rangle_{\phi,-}f_{-}\right),\label{eeq16b}\\
\frac{dJ'}{dt'}&=&M\left(\langle H\rangle_{\phi,+}f_{+}+\langle H\rangle_{\phi,-}f_{-}\right)-{J'}+\Gamma'_{H_{2}}\overline{\mc\theta}(f_{+}-f_{-}),\label{eeq17b}
\ena
where $\Gamma'_{H_{2}}$ is the dimensionless H$_{2}$ torques,
\bea
M=\frac{\gamma \overline{\lambda} u_{rad}a^{2} t_{gas}}{2 J_{th}}\label{m},
\ena
and  $f_{+}, f_{-}$ are the fraction of the time interval $dt$ the grain stays in the positive and negative flipping states, respectively (see Appendix D); $\langle F\rangle_{\phi(+,-)}$ and $\langle H\rangle_{\phi(+,-)}$ correspond to positive and negative flipping states (see HL08a).\footnote{The positive and negative flipping states are defined based on the initial directions of $\ba_{1}$ with respect to $\bJ$. The positive flipping state corresponds to $\ba_{1}$ parallel to $\bJ$ and the negative flipping state corresponds to $\ba_{1}$ antiparallel to $\bJ$.} 

\begin{figure}
\includegraphics[width=0.5\textwidth]{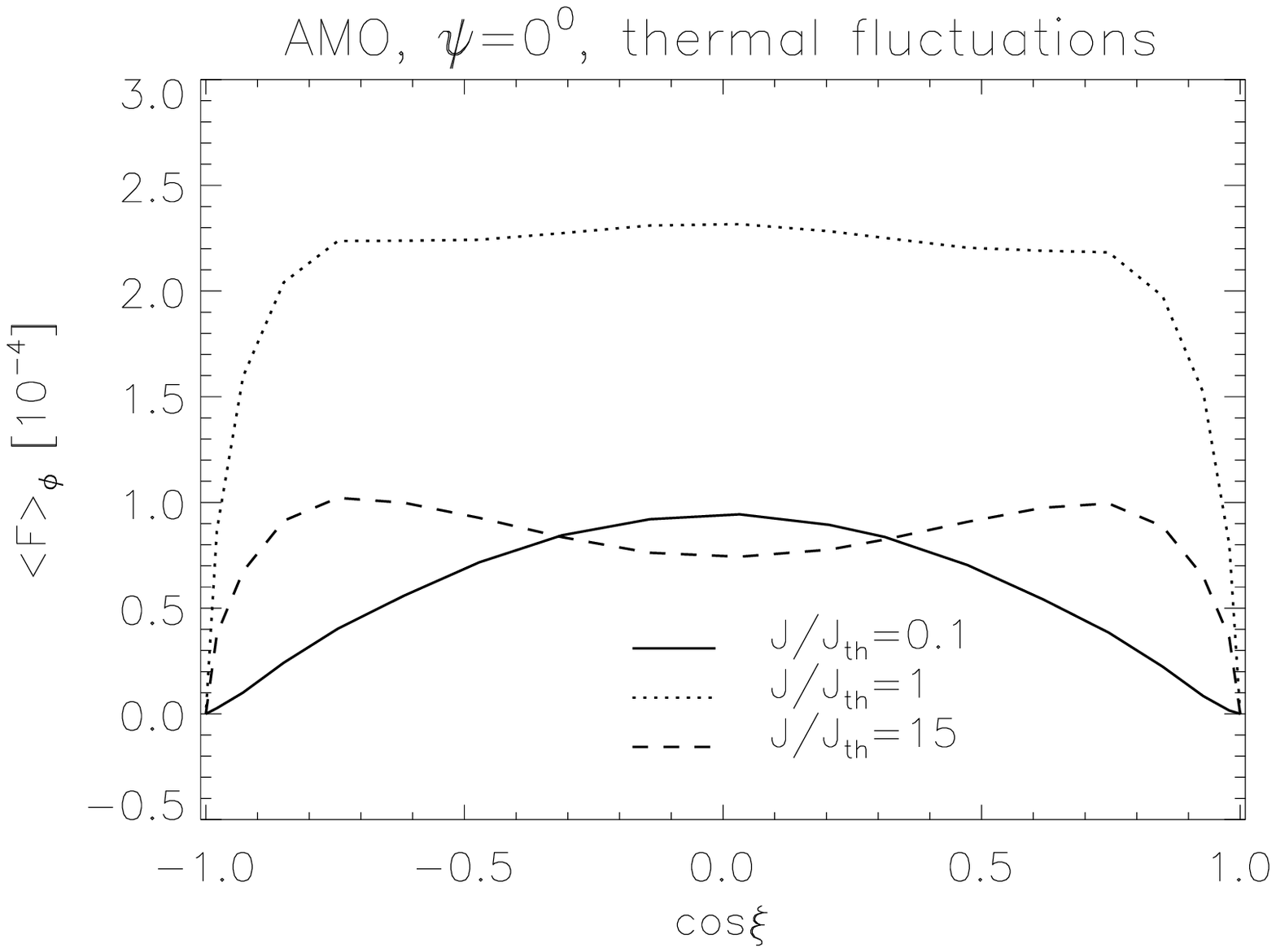}
\includegraphics[width=0.5\textwidth]{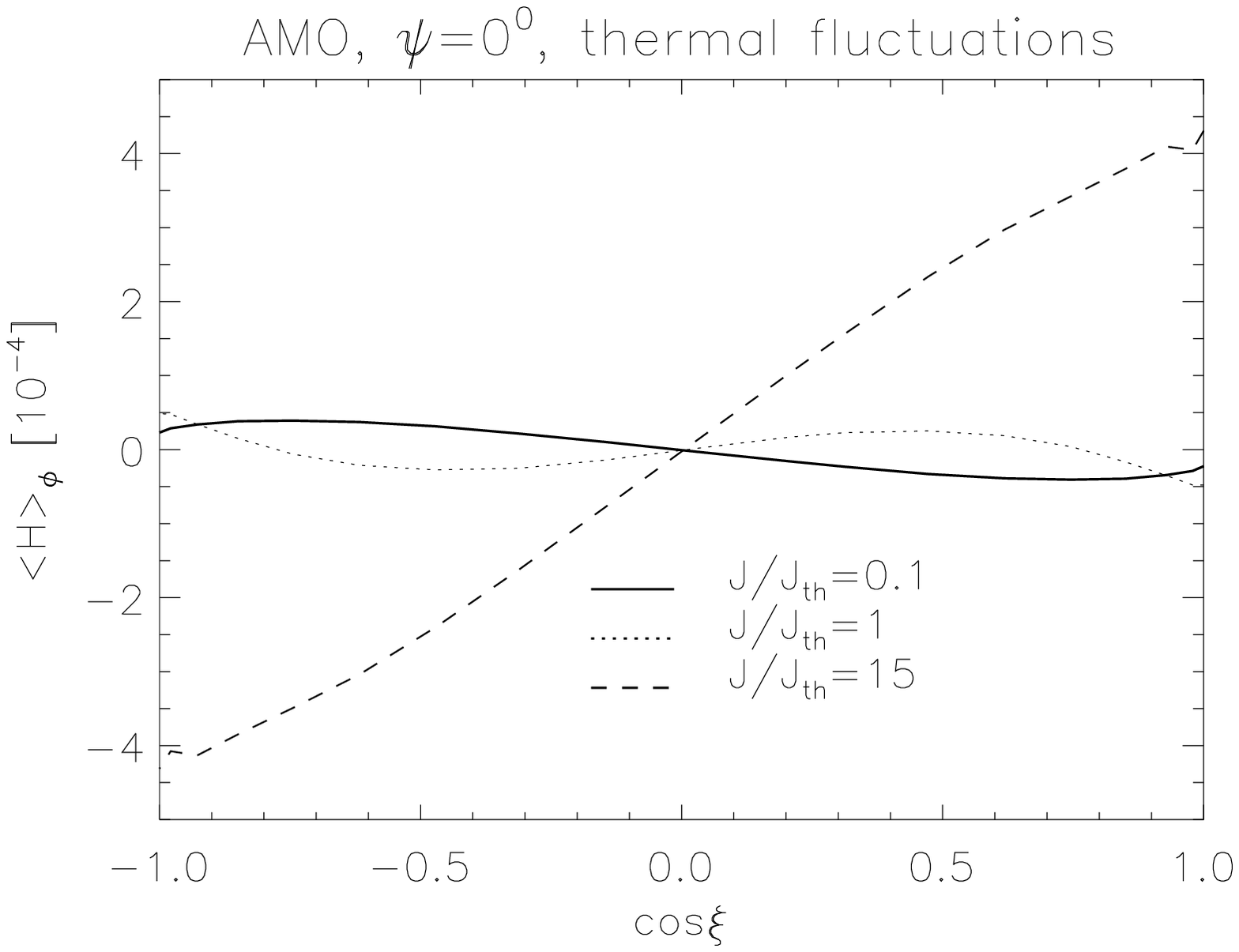}
\caption{Aligning ({\it upper}) and spin-up ({\it lower}) components of RATs for the ellipsoidal AMO with $q^{max}=1.2$ and ratio of inertia moments $I_{1}:I_{2}:I_{3}= 1.745:1.61:0.876$ and for light direction $\psi=0^{\circ}$, obtained by averaging over thermal fluctuations for different values of angular momentum $J$. Lower $J$ corresponds to stronger thermal fluctuations and the magnitude of RATs decrease. Functional forms of RATs also change as $J$ decreasing, but stationary points $\xi=0$ and $\pi$ remain unchanged.}
\label{f10}
\end{figure}

We adopt the RAT efficiency ${\bf Q}_{\Gamma}$ from the AMO with the ratio of two first components of torques $q^{max}=1.2$ (see Appendix C). We first average RATs over torque-free motion, and thermal fluctuations for the irregular grain with  ratio of inertia moments $I_{1}:I_{2}:I_{3}=1.745:1.61:0.876$ (see WD03; HL08a) to get  $\langle Q_{e_{j}}\rangle$. Then, we calculate aligning and spin-up components of RATs using equations (C13) and (C15) where $Q_{e_{j}}, j=1-3$ are replaced by $\langle Q_{e_{j}}\rangle$. The resulting aligning and spin-up components of RATs after averaging over precession angle $\psi$ are shown in Figure \ref{f10} for three values of $J$ and $\psi=0^{\circ}$. For this setting of AMO, the two stationary points $\mc\xi=\pm 1$ are expected (see the upper panel). For the spin-up component, $\langle H\rangle_{\phi}$ at these stationary points changes the sign to the opposite when $J\rightarrow 1$. As a result, we expect low-J attractor points in trajectory maps for this alignment (see HL08a).
 
\subsection{Trajectory maps}
Using RATs from Figure \ref{f10}, we solve equations of motion for $J$ and $\xi$. 
We use initial conditions $J_{0}=J_{th}$ and $\xi_{0}$ is generated from a uniform distribution of the angle between $\bJ$ and $\bB$. For long-lived H$_{2}$ torques, we assume $t'_{L}=10^{2}$ and a constant time step $\delta t'=10^{-3}$. We assume again that  H$_{2}$ torques are parallel to the axis of maximal inertia for positive flipping state.

\begin{figure}
\includegraphics[width=0.49\textwidth]{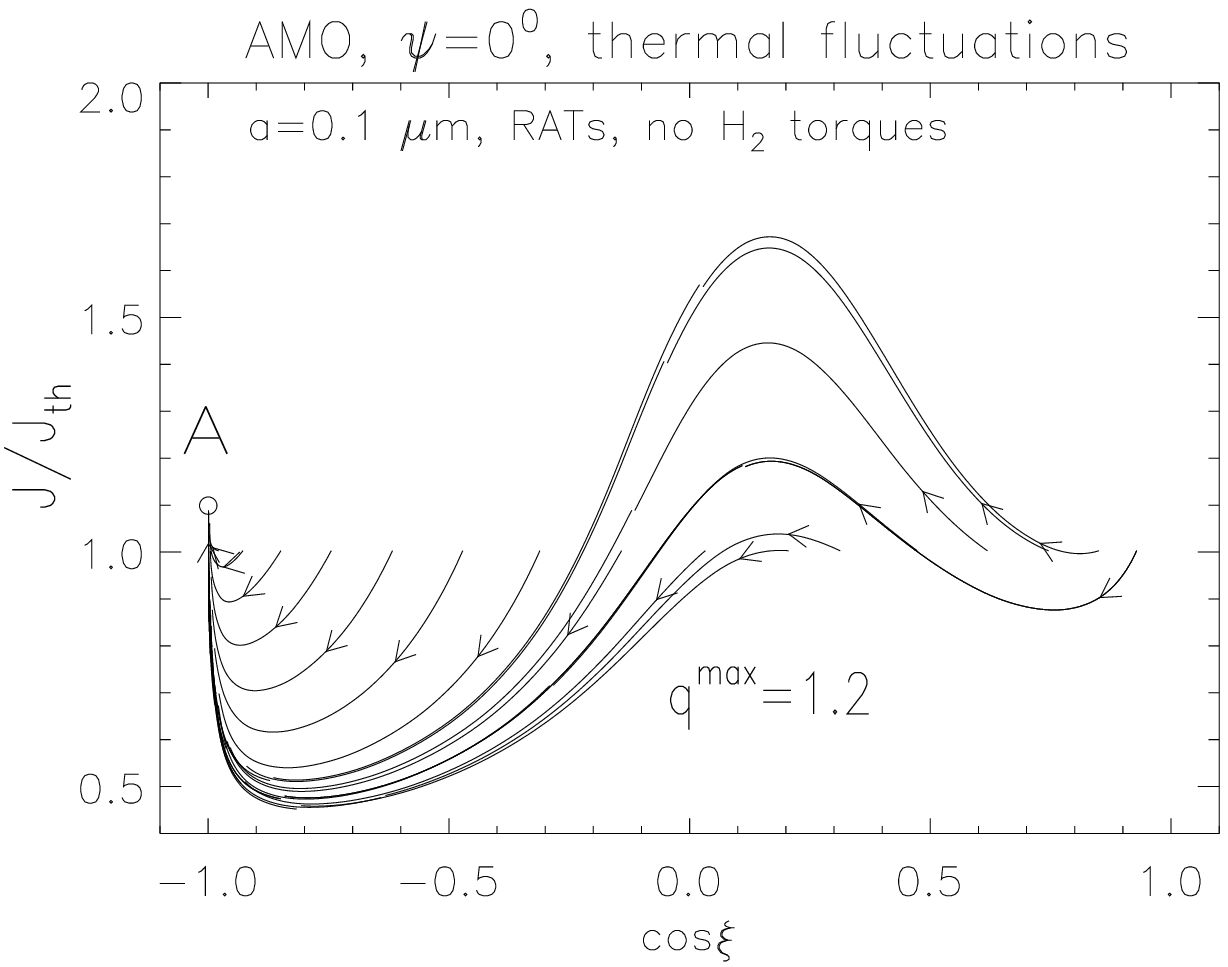}
\includegraphics[width=0.5\textwidth]{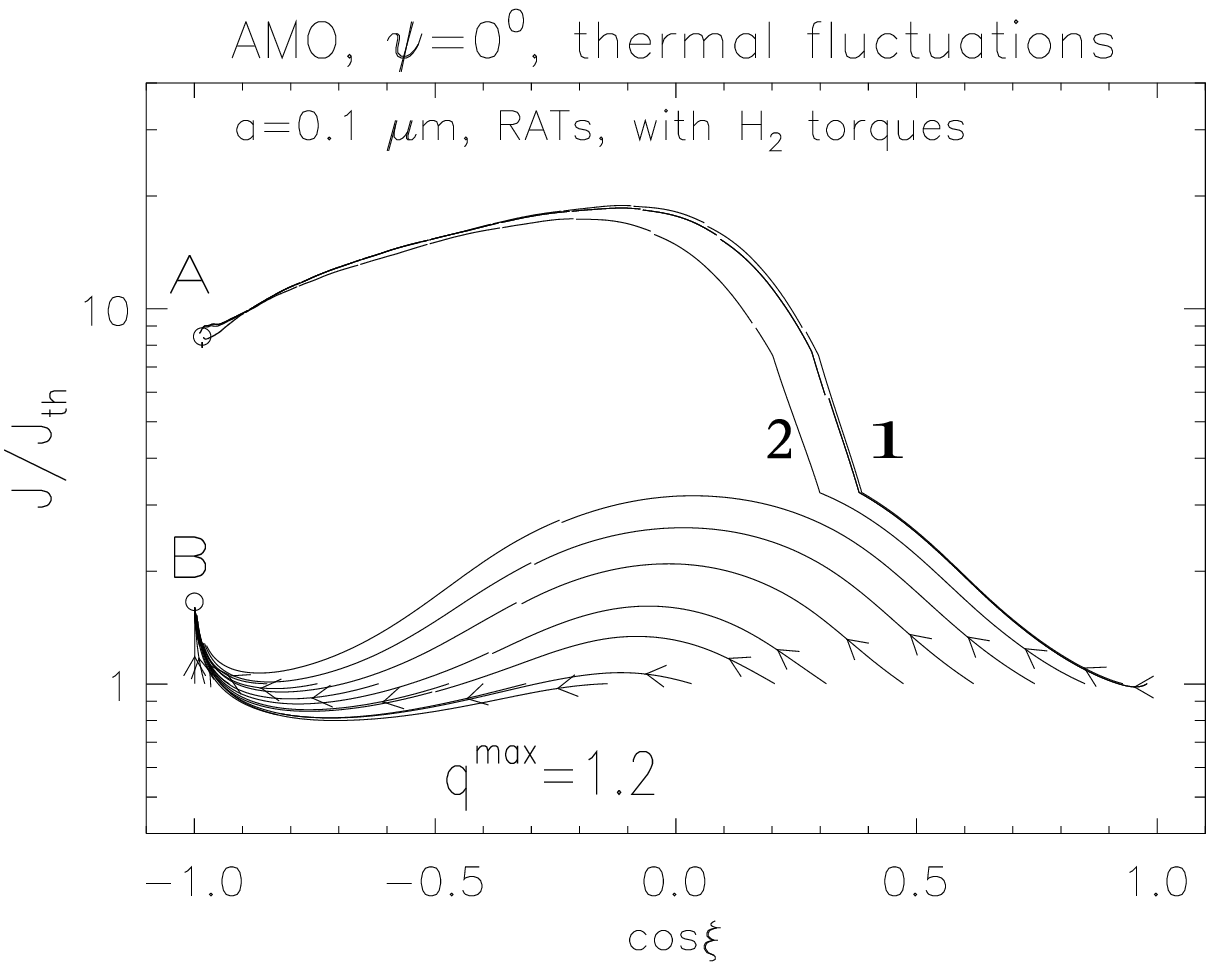}
\caption{Phase trajectory maps for the alignment in the diffuse ISM of a $0.1\mu m$ grain for light direction $\psi=0^{\circ}$, in the absence of H$_{2}$ torques ({\it upper panel}) and in the presence of long-lived H$_{2}$ torques with $\alpha=5\times 10^{14}$ cm$^{-2}$ ({\it lower panel}). RATs align grains at a low-$J$ attractor point B seen in both panels, but the presence of long-lived H$_{2}$ torques gives rise to the new high attractor point (A) corresponding to suprathermal rotation in the lower panel. The AMO with torques ratio $q^{max}=1.2$ and the ISRF are adopted.}
\label{f11}
\end{figure}

The trajectory maps with RATs from Figure \ref{f10} are presented in Figure \ref{f11}. The upper panel represents the alignment by RATs only, and the lower panel represents the alignment by both RATs and the long-lived H$_{2}$ torques.

The upper panel of Figure \ref{f11} shows the trajectory maps for the alignment by RATs only. For this case, all grains with different initial angles $\xi$ are driven to a low-J attractor point A.
In the presence of H$_{2}$ pinwheel torques, the lower panel of Figure \ref{f11} shows that such long-lived torques can produce a new attractor point of high angular momentum (point B) with $J/J_{th}\sim 9$. The alignment angle corresponds to the perfect alignment of $\bJ$ with respect to $\bB$ (i.e. $\mc\xi=-1$).

The appearance of a new high-J attractor point at $\mc\xi=-1$ in the presence of long-lived H$_{2}$ torques can be easily understood. First, some grains with initial direction in the vicinity of $\mc\xi=0$ are spun up by positive spin-up torques (see Fig. \ref{f10}) so that flipping rate decreases. Thus, pinwheel torques become more efficient and act to increase $J$. When $J\ge 3J_{th}$, the flipping decreases substantially and the grains are able to escape from the thermal trapping (grains with trajectories 1 and 2). At the same time, the aligning component of RATs drives the grains to the stationary point $\mc\xi=-1$ as in the absence of pinwheel torques. Finally, these grains are stably aligned at the high$-J$ attractor point. The rest of grains are aligned at low-$J$ points.

In fact, as seen in Figure \ref{f10}, the aligning torque has two stationary points at $\mc\xi_{s}=\pm 1$ for $J\gg J_{th}$. In the absence of H$_{2}$ torques, this stationary point is a low-J attractor point because of the spin-up component of RATs $\langle H\rangle_{\phi}(\xi_{s}=\pi) <0$. In the presence of positive H$_{2}$ pinwheel torques, which are assumed to be larger than RATs, we have the total spin-up torque $\langle H\rangle_{tot}=M\langle H\rangle_{\phi}+\Gamma_{z}^{'b}>0$. Thus, $\left.\langle H\rangle_{tot}\frac{ d\langle F\rangle_{\phi}}{d\xi}\right|_{\xi_{s}=\pi}<0$ because $\left.\frac{ d\langle F\rangle_{\phi}}{d\xi}\right|_{\xi_{s}=\pi}<0$ (see Fig. \ref{f10} ). This indicates that the stationary point $\mc\xi_{s}=-1$ becomes an attractor point.

We stress that the role of H$_{2}$ torques in producing the perfect alignment of $\bJ$ with $\bB$ in Figure \ref{f11} does not involve the paramagnetic dissipation. Instead, the stationary points are produced by the RATs (see LH07a). 

\subsection{Increase of $J$ at low attractor points}
In the absence of pinwheel torques, the value of $J$ at low-J attractor points (hereafter $J_{low}$) is constrained by the internal relaxation as well as the negative spin-up component of RATs.
The presence of pinwheel torques increases the value $J'_{low}$ as (see eq. \ref{eeq17b})
\bea
J'_{low}(\psi)&=&M\langle H\rangle_{\phi}+\Gamma_{z}^{'b}(f_{+}-f_{-}),\nonumber\\
&=&-\epsilon_{J'}\epsilon_{\psi} J_{max}^{'RAT}+J_{max}^{'H_{2}}.\overline{\mc\theta}(f_{+}-f_{-}),\label{jlow}
\ena
where we used the fact that $\langle H\rangle_{\phi,+}\sim \langle H\rangle_{\phi,-}$ and $f_{+}+f_{-}=1$ in eliminating the term $f_{+},f_{-}$ associated with $\langle H\rangle_{\phi}$, $\epsilon_{J'}$, evaluated at $J'=J'_{low}$, is the factor by which $\langle H\rangle_{\phi}$ is reduced by averaging over thermal fluctuations at fixed $J'$, $\epsilon_{\psi}$ is a function of $\psi$ describing the decrease of spin-up torque with $\psi$. At $J=J_{th}$, we have $\epsilon_{J} \sim 0.1$ (see HL08a). From our calculation of RATs using DDSCAT for the shape 1, we found that $\epsilon_{\psi}=1$ for $\psi=0^{\circ}$ and $\sim 0.25$ for $\psi=70^{\circ}$ for a $0.2\mu$m grain (see also HL08b).

From equation (\ref{jlow}) it is obvious that for fast flipping, i.e., $f_{+}-f_{-}=0$, the pinwheel torques play a marginal role in spinning up grains. However, our calculations in \S 2 show that the flipping gets slower for grains larger than $10^{-5} cm$, so the pinwheel torques can have an important effect for these grains.

\begin{figure}
\includegraphics[width=0.5\textwidth]{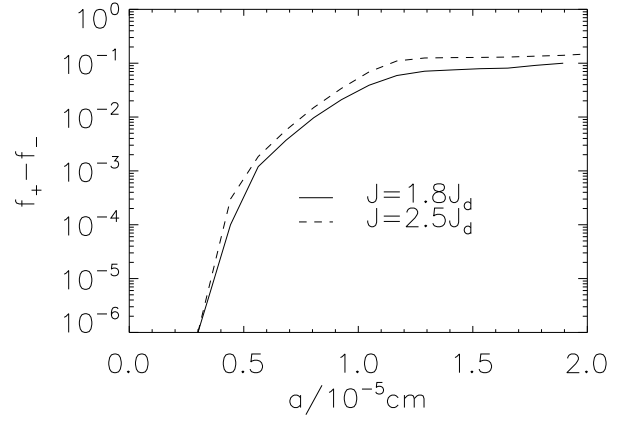}
\caption{Difference between the fraction of the time the grain spends in the positive flipping state and the fraction of time the grain spends in the negative flipping state as a function of the grain size for $J=1.8 J_{d}$ and $2.5 J_{d}$ assuming that the grain is in positive flipping state at the beginning of the time interval with time step $dt=10^{-3}t_{gas}$. $f_{+}-f_{-} \le 10^{-6}$ for $a < 0.3 \times 10^{-5}$cm.}
\label{fs}
\end{figure}

Using equation (\ref{ap114}) and adopting the time step $dt=10^{-3}t_{gas}$, we can estimate $f_{+}-f_{-}$ for different $J$ and grain sizes assuming the grain is in positive flipping state at the beginning of the time interval. Figure \ref{fs} shows that for grain smaller than $\sim 3\times 10^{-5}$cm, we have $f_{+}-f_{-}\approx 0$. For larger grain $a=10^{-5}$cm, $f_{+}-f_{-}=0.04$ for $J=J_{th}=1.8J_{d}$. Combining equations (\ref{eq12}), (\ref{rat1}) and (\ref{jlow}) with Figure \ref{fs}, we can obtain $J'_{low}$ as a function of $\alpha$ and grain size.

We note that the mean flipping time increases with $J$, therefore, at some value of $J$ (i.e., $J\ge 5$), the grain does not flip. Thus, $f_{+}-f_{-}=1$, and the low-J attractor point transfers to high-J attractor point (see \S 4.2). 

\begin{figure}
\includegraphics[width=0.5\textwidth]{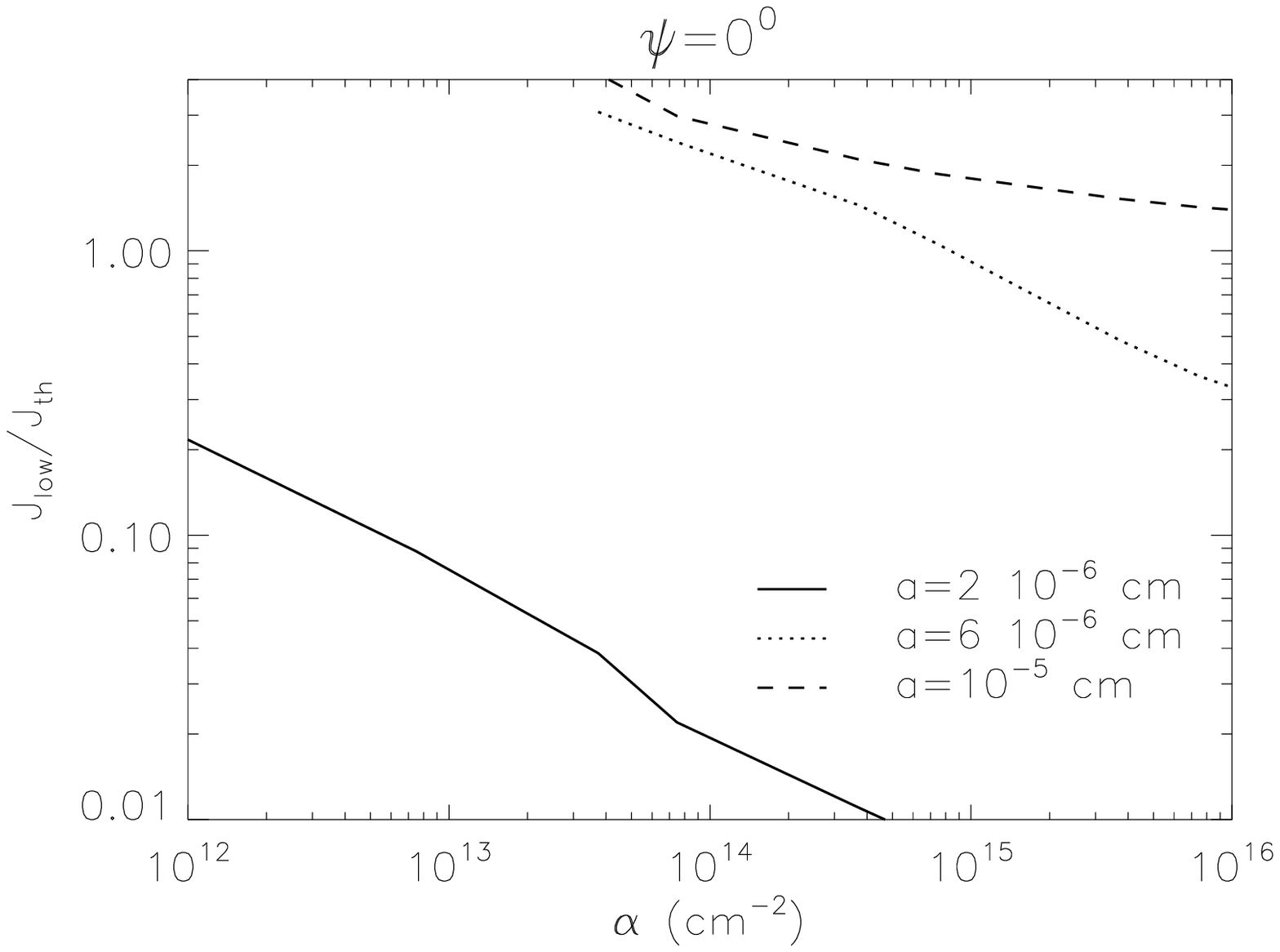}
\includegraphics[width=0.5\textwidth]{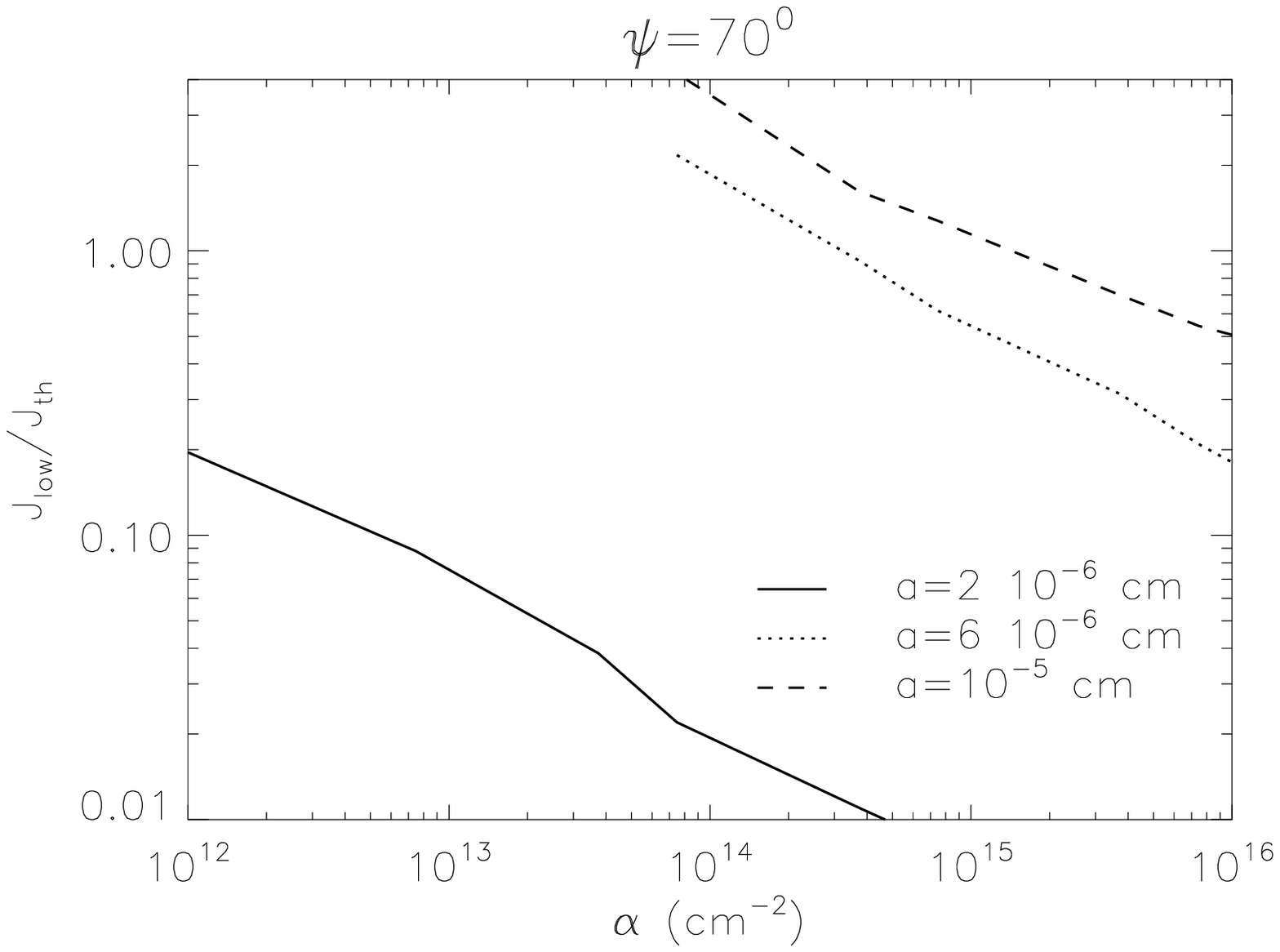}
\caption{Value of angular momentum at low attractor points as a function of the active site density $\alpha$ for different grain sizes and two light directions $\psi=0^{\circ}$ ({\it upper panel}) and $\psi=70^{\circ}$ ({\it lower panel}). The AMO similar to Fig. \ref{f11} is adopted. }
\label{f12}
\end{figure}

To get the exact value of $J'_{low}$, we solve the equations of motion and construct the trajectory maps. 
Figure \ref{f12} represents the obtained results  using the AMO with torques ratio $q^{max}=1.2$ for two radiation directions $\psi=0^{\circ}$ ({\it upper panel}) and $70^{\circ}$ ({\it lower panel}) and for three values of grain size ($a=2\times 10^{-6}$, $6\times 10^{-6} $ and $10^{-5}$ cm). For both directions, the value $J_{low}$ increases fast with the decrease of $\alpha$, i.e., the value of H$_{2}$ torques increases. However, for the grain with $a\le 2\times 10^{-6}$ cm, $J_{low}$ can not exceed the thermal value even with the strongest H$_{2}$ torques. This arises from the fast flipping of such small grains that decreases significantly the effect of H$_{2}$ torques. For the former case, $J_{low}< J_{th}$ for all possible magnitudes of pinwheel torques. This result is consistent with our calculation in the preceding section.

The above results are obtained using the default ellipsoidal AMO. However, since the H$_{2}$ torques depend  weakly upon the shape of grains, we expect the results are  valid for irregular grains.

\subsection{Effect of pinwheel torques on the alignment with high-$J$}
 Now let us consider the effect of pinwheel torques on the RAT alignment in the ISM for a large grain $a=0.2\mu$m for which the flipping is inefficient, and the torque ratio $q^{max}=0.78$ to get the alignment with high-$J$ attractor point for the radiation direction $\psi=70^{\circ}$ (see LH07a). We assume that the pinwheel torques are parallel to $\ba_{1}$ axis and have a magnitude with $\alpha=10^{16}$cm$^{-2}$.

\begin{figure}
\includegraphics[width=0.5\textwidth]{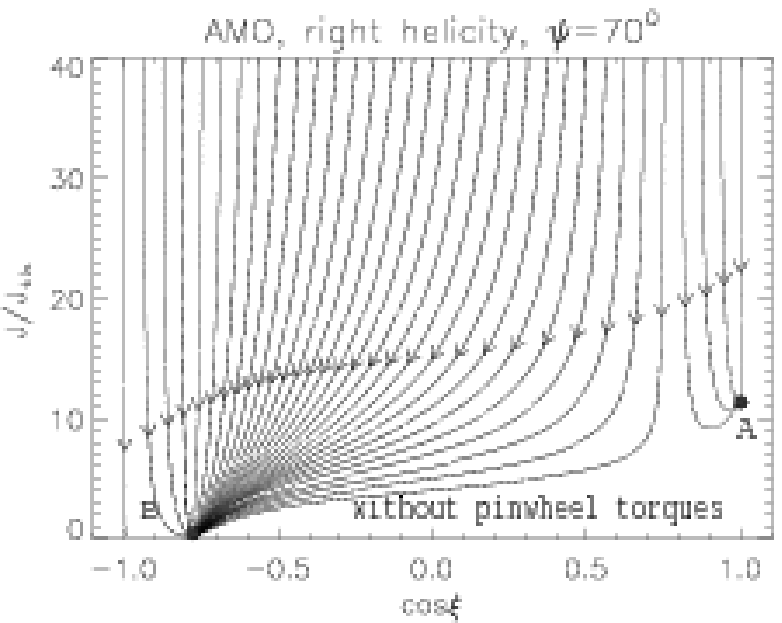}
\includegraphics[width=0.5\textwidth]{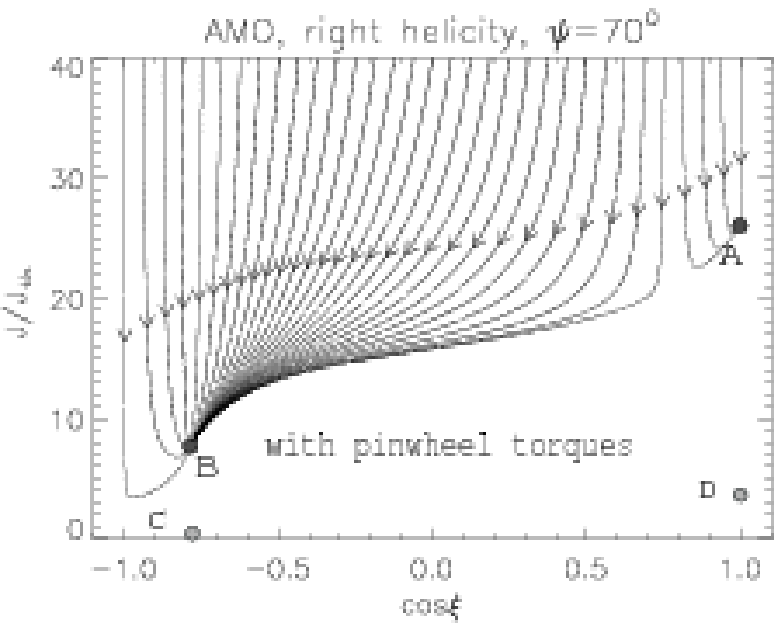}
\caption{Trajectory maps of the RAT alignment for a grain size $a=0.2\mu m$ without and with pinwheel torques (upper and lower panels) for $\alpha=10^{16}$~cm$^{-2}$. The presence of pinwheel torques parallel to $\ba_{1}$ axis results in the high-J attractor point B, but does not change the alignment angle. C and D are expected attractor points when pinwheel torques are anti-parallel to $\ba_{1}$ and the amplitude of pinwheel torques is smaller than that of RATs.}
\label{f12b}
\end{figure}

Figure \ref{f12b} shows the trajectory maps in the absence and presence of pinwheel torques. In the first case, RATs align grains with a high-$J$ attractor point A and a low-$J$ attractor point B. In the second case, the presence of pinwheel torques parallel to $\ba_{1}$ axis increases the angular momentum of the attractor points A and B. The low-$J$ attractor point becomes the high-$J$ attractor point, but the alignment angle $\mc\xi=-0.8$ does not change.

When the amplitude of pinwheel torques is less than that of RATs, the pinwheel torques can provide a new type of attractor point. Indeed, at a high attractor point the pinwheel torques may act opposite to RATs, moving point A to the position D. The point C is a regular crossover point with the crossovers assisted both by RATs and pinwheel torques. As the crossover happens, the grain flips (see Spitzer \& McGlynn 1979, Lazarian \& Draine 1997, 1999a) and the RATs and the pinwheel torques start acting in the opposite directions. This creates a new uplifted low attractor point similar to B, but corresponding to the opposite state of rotation. The individual realizations of pinwheel torques make the trajectory maps not symmetric, but if we consider an ensemble of grains with equal number of positive and negative pinwheel torques the trajectories for AMO get symmetric and we can show only an upper part of them as we do in Figure \ref{f12b}. 

The most important effect of the strong pinwheel torques is their possible uplifting of the low attractor point. This should increase grain alignment in the case when the trajectory maps do not have high attractor points. The latter situation corresponds to a substantial portion of the parameter space (see LH07a).  

\subsection{Role of RATs in spinning-up grains}
In \S 5.2 we found that positive RATs can support the pinwheel torques to bring grains to high-$J$ attractor points because they are not fixed in the grain body axes. Now let us estimate the value of $J$ that spin-up RATs can induce the grain for a given light direction $\psi$.

The spin-up component of RATs for a light direction $\psi=70^{\circ}$ is shown Figure \ref{f10}. For $J\ge J_{th}>J_{d}$, $\langle H\rangle_{\phi}>0$ for $\mc\xi>0$, and $\langle H\rangle_{\phi}<0$ otherwise (see also the lower panel of Fig.\ref{f10}). The magnitude of spin-up torque at the angle $\xi=0$ is given by equation (\ref{eq9c}) when the radiation is parallel to the magnetic field. For an arbitrary radiation direction $\psi$, the magnitude is decreased by a factor of $\epsilon_{\psi}$. We can estimate the maximal angular momentum spun-up by RATs when radiation makes an angle $\psi$ with $\bB$ as followings
\bea
J'_{max}(\psi)=M\langle H\rangle_{\phi}(\psi,\xi=0)=\epsilon_{J'} \epsilon_{\psi}J_{max}^{'RAT},\nonumber\\
\approx8\times 10^{3}\frac{\gamma}{0.1}\hat{\rho}^{1/2}a_{-5}^{1/2}  \left(\frac{\bar\lambda}{1.2~\mu m}\right)^{\eta+1}\left(\frac{12}{a_{-5}}\right)^{\eta}\epsilon_{J'} \epsilon_{\psi},\label{eqj1}
\ena
where  $J_{max}^{'RAT}$ is given by equation (\ref{rat1}).

For $\lambda> 1.8 a$, $\eta=-3$, we obtain
\bea
J'_{max}(\psi)&\approx 4.6 a_{-5}^{7/2}\epsilon_{J'} \epsilon_{\psi},\label{eqj}
\ena
It can be seen that $J_{max}$ increases rapidly with the grain size as a function of $a_{5}^{7/2}$. For a grain size $a_{-5}=2$, we can obtain $J'_{max}\approx 12$ for $\psi=70^{\circ}$ with $\epsilon_{\psi}=0.25$ and $\epsilon_{J'}=1$.\footnote{We note that as $J'$ increases, the factor $\epsilon_{J'}$ in equation (\ref{eqj}), which describes the decreases of the angular momentum due to the thermal fluctuations, increases, and achieve the order of unity for $J'\ge 10$ (see HL08a)} When grains are spun-up by RATs to $J_{max}\gg J_{th}$, flipping stops, and H$_{2}$ torques become efficient. From equation (\ref{eqj}) it can be seen that for grains $a_{-5}\ge 1$, $J'_{max}\ge 1 $ for every $\psi$. This indicates that RATs for the ISRF  are able to spin up large grains to suprathermal rotation, and thus they are not thermally trapped.

\section{Discussion and Summary}

\subsection{Grains are flipping and get trapped}

New elements of grain dynamics, namely, thermal flipping and thermal trapping were reported in LD99a. The validity of these findings was challenged in Weingartner (2008), who found that grains can not experience thermal flipping as a result of internal relaxation. In the present work we showed that although internal relaxation can not result in grain thermal flipping, but in the presence of impulses due to H$_{2}$ formation and gas bombardment, grains can flip, and get trapped. The impulses play the role of an engine to bring grains through the boundary $\ba_{1}\perp\bJ$.

The thermal trapping is an essential process that, according to LD99a, explains why, in accordance with observations (see Kim \& Martin 1995), small grains are not aligned by the Purcell (1979) paramagnetic relaxation process. The alternative explanations to this observational fact (see Lazarian 1995) are much less appealing. The sizes of the thermally trapped grains are not different from those which can be
obtained using the LD99a treatment, but smaller than those discussed in LD99b. As a result, pinwheel torques are more important for interstellar grains than one could infer from LD99b. 

\subsection{Extending theory of crossovers}

Crossovers are times when grains subject to pinwheel torques slow down, then flip and get accelerating. The original theory of crossovers was suggested by Spitzer \& McGlynn (1979). Lazarian \& Draine (1997) improved the theory by accounting for the value of the residual thermal angular momentum,
associated with thermal wobbling of the grain (see Lazarian 1994). LD99b showed that with nuclear relaxation taken into account the theory for regular crossovers in the spirit of the aforementioned works is applicable only for grains larger than $\sim 10^{-4}$~cm, which excludes most of grains in diffuse interstellar gas. LD99a, instead, introduced the concepts of thermal flipping and trapping for smaller grains.

The present paper shows that the flipping and trapped grains are smaller than those discussed in earlier papers. Therefore, in the absence of RATs, there is a range of grains for which the crossovers happen over the time scale larger than the internal relaxation time. In this situation the thermal value of the angular momentum can decrease below  $J_d$ a result of the action of the pinwheel torques. This increases
the randomization of grains during crossovers for the range of sizes larger than the flipping size.   

\subsection{How common is suprathermal rotation?}

Suprathermal, i.e. much faster than thermal, rotation was assumed to be default for most of the interstellar grains after the Purcell (1979) study. Studies of RATs in
DW96, DW97 only made this point stronger. However, later works that reported thermal trapping of grains (LD99a, LD99b) and low-J attractor points (Weingartner \& Draine 2003, LH07a) made one wonder whether most grains rotate thermally. The study in HL08a revealed that in the presence of gaseous bombardment or
other randomization processes, the grains tend to diffuse from the low-J to high-J attractor point, provided that such high-J attractor point coexist with the low-J one.
According to the parameter space study in LH07a (see Fig. 24 in LH07a) this corresponded to a fraction of grains that could vary depending on the angle between the radiation direction and the magnetic 
field, as well as on the $q^{max}$ ratio. 

In the present study we showed that for sufficiently large grains the pinwheel torques, provided that they are comparable with or stronger than the RATs, can induce suprathermal rotation of grains.  
Interestingly enough, this can happen even in the situations when only low-$J$ attractor points exist. Therefore, the presence of pinwheel torques can increase the percentage of the grains
rotating suprathermally. However, it is worth noting, that for a substantial
portion of the parameter space radiative torques act against pinwheel torques attempting to decrease the grain rotation rate.

\subsection{Implications to polarization modeling and magnetic field diagnostics}

A quantitative theory of grain alignment is very important to polarization simulations. A number of works dealing with polarization simulations assume the perfect alignment of the grain axis with respect to the magnetic field (see CL07; Bethell et al. 2007; Pelkonen et al. 2007; Falceta-Goncalves, Lazarian \& Kowal 2008). A detailed comparison between the polarization efficiency predicted by the theory of grain alignment and observations of dense cloud in Whittet et al. (2008) shows that a high degree of alignment is required to explain the observed polarization degree. How can this be accomplished?

In Lazarian \& Hoang (2008) we proposed one way of increasing the degree of grain alignment, namely, we noticed that in the situation when grains have superparamagnetic inclusions the high-$J$ attractor points
are present for all the angles between the radiation direction and the magnetic field, as well as on the $q^{max}$ ratio. Thus the gaseous bombardment is expected to move all grains to the
high attractor points, making the alignment perfect. In the present paper we show an alternative way of increasing the degree of grain alignment. In particular, we show that RATs plus pinwheel torques can enhance the degree of alignment by increasing the value of angular momentum at the low-$J$ attractor point. However, the circumstances
for which the pinwheel torques affect the RAT alignment are rather restrictive: the pinwheel torques are required to be comparable or stronger than RATs.

With these theoretical results one can attempt to distinguish between the two alternatives above and simultaneously get insight into the grain composition. First of all, more studies similar to those done by Whittet et al. (2008) are required to make sure that the high alignment is not a result of favorable illumination geometry. Alternatively, one may make measurements in situations where the illumination geometry is known. Comparing these results with the analytical and numerical predictions (see LH07a, HL08a) one can establish with higher certainty that the grain alignment is enhanced compared to what is expected from
ordinary paramagnetic grains subject to RATs in the absence of the pinwheel torques. In parallel, one can search for the correlation of grain alignment enhancement with the environments where we expect the higher amplitudes of the pinwheel torques. If such an enhancement is present, it will indicate that grains are {\it not} superparamagnetic. If they were superparamagnetic, we would expect the perfect alignment for typical conditions of the diffuse ISM, which one cannot be improved further by the action of the pinwheel torques. No correlation, but high degrees of alignment mean, on the contrary, the existence of superparamagnetic grains.

One can conclude by observing an interesting cycle in the development of the grain alignment theory. The initial theory by Davis-Greenstein (1951) appealed to paramagnetic dissipation in ordinary paramagnetic grains. The mechanism was shown not to be sufficiently strong, however. Then to enhance the efficiency of the initial Davis-Greenstein process both superparamagnetism  (Jones \& Spitzer 1967) and
pinwheel torques (Purcell 1979) were appealed to. At the moment we may face the situation that the RATs are not strong enough to explain the alignment and again we appeal to superparamagnetism (Lazarian \& Hoang 2008) and pinwheel torques (this paper). Nevertheless, there is a substantial difference between the two situations. The paramagnetic relaxation in the diffuse interstellar medium could provide a few
percent alignment at most. The RAT alignment provides the degrees of alignment about 20\% even in the cases of unfavorable directions of illumination (see HL08a). In addition, while the Davis-Greenstein process
favored the alignment of small grains, the RAT alignment, in accordance with observations, favors the alignment of large grains.

\subsection{Summary}
Our main results are summarized as follows:
\begin{itemize}

\item We extended the definition of the pinwheel torques to include three new processes, namely, radiative torques arising from grain infrared emission, torques due to interactions of
grain with electrons in plasma, as well as, torques arising from gas flow interacting with a helical grain. We proposed to consider the radiative torques arising from the action of 
isotroptic radiation on a grain as pinwheel torques. All these torques we identified as long-lived, as their life-time is expected
to be longer than the typical grain rotational damping time. We used H$_2$ torques as a proxy of for general pinwheel torques, but, unlike earlier studies, we allowed the pinwheel torques to be both short-lived and long-lived. We found that in the ISM, the pinwheel torques due to infrared emission is in general weaker than that due to H$_{2}$ formation, and RATs.

\item We proved that flipping and trapping of astrophysical dust grains is the active and important process. We did this by augmenting the torques arising from internal relaxation by the stochastic torques that inevitably accompany the action of pinwheel torques and gas bombardment. . We calculated the thermal flipping induced by the Barnett, nuclear relaxations, as well, as internal relaxation in superparamagnetic grains in the presence of impulses due to H$_{2}$ formation and gas bombardment. Adopting the revised diffusion coefficients for internal relaxation from Weingartner (2008), we found that flipping is fast for small grains, and increases when superparamagnetic inclusions are accounted for. We obtained the critical size of flipping and trapping size of the thermally trapped grains for the ISM. 

\item The most important result of this study is related to the increase of the expected degree of grain alignment when dust grains are subject to both RATs and pinwheel torques. This increase stems from
the increase by pinwheel torques of the value of the angular momentum at the low-$J$ attractor point. The increase of angular momentum decreases the wobbling of grains at the low-$J$ attractor point and
therefore increase the degree of internal alignment as the internal alignment (of $\ba_{1}$ with $\bJ$) is nearly perfect when the grain rotates with rates much larger than the thermal rotation rate. The joint pinwheel plus RATs alignment is different from the Purcell (1979) alignment. In the latter mechanism the alignment is due to paramagnetic dissipation, which induces torques that are very weak compared to RATs. 
 \end{itemize}

\acknowledgements
We thank the referee, Bruce Draine, for helpful comments that significantly improve our paper.
 We thank Wayne Roberge and K.E. Saavik Ford for sharing with us their results from Roberge \& Ford (2000). The work was supported by the NSF Center for Magnetic Self-Organization in Laboratory and Astrophysical
Plasmas and NSF grant AST 0507164.

\appendix

\section{Diffusion coefficients}

LR97 derived the diffusion coefficients for the Barnett effect by assuming that in thermal equilibrium, the probability current
\bea
S=A(\theta)f_{TE}(\theta,T_{d})-\frac{1}{2}\frac{\partial}{\partial \theta}\left( B(\theta)f_{TE}(\theta,T_{d})\right).
\ena
with $f_{TE}((\theta,T_{d})$ the Boltzman distribution function, is equal to zero.

Weingartner (2008) used instead of angle $\theta$ a variable $q=\frac{2I_{1}E}{J^{2}}$, where  for an oblate spheroid
$E=\frac{J^{2}}{2I_{1}}\left(1+[h-1]\ms^{2}\theta\right)$ and found diffusion coefficients different from those in LR97, namely
\bea
A(q)&=&-\frac{(q-1)(h-q)}{t_{int}(J)},\\
B(q)&=&\frac{1}{\zeta^{2}t_{int}(J)}\left[ 3+2\zeta(q-1)\right](h-q)\nonumber\\
&&+C(\zeta)(h-q)^{1/2}\mbox{exp}(\zeta q)-\zeta^{-1/2}\left[ 3+2\zeta(h-1)\right]\nonumber\\
&&\times(h-q)^{1/2}\mbox{exp}[-\zeta(h-q)]\int_{0}^{\sqrt{\zeta(h-q)}} \mbox{exp}(x^{2})dx,\nonumber\\
\ena
where $h=I_{1}/I_{2}, \zeta=(J/J_{d})^{2}$, $B(q=h)=0$. We would claim that the main difference between these coefficients and those in LR97 is that both 
$A$ and $B$ coefficient get zero when the angle is equal $\theta$ is equal to 90 degrees.

Dealing with oblate grains, we prefer, however, to study the dynamics in terms of $\theta$. The corresponding Ito's formulae (see Gardiner 1983) 
for the transformation of variables in the stochastic calculus provide us with:
\bea
d\theta=\left(A(q)\frac{\partial \theta}{\partial q}+\frac{B(q)}{2}\frac{\partial^{2} \theta}{\partial q^{2}}\right)dt+\sqrt{B(q)}\frac{\partial \theta}{\partial q}dw,
\ena
where $dw$ is a random variable with $\langle dw^{2}\rangle=dt$.
Thus, using the equivalence of Langevin equation and Fokker-Planck equations, we obtain
\bea
A(\theta)&=&A(q)\frac{\partial \theta}{\partial q}+\frac{B(q)}{2}\frac{\partial^{2} \theta}{\partial q^{2}}=\frac{-(h-1)}{2t_{int}(J)}\ms\theta\mc\theta+B(q)\frac{-\mc2\theta}{(h-1)^2\mbox{sin}^{3}2\theta},\label{abar}\\
B(\theta)&=&B(q)\left(\frac{\partial \theta}{\partial q}\right)^{2},\label{bbar}
\ena
where we used the relation $q=1+(h-1)\mbox{sin}^{2}\theta$ to calculate first and second derivative of $\theta$ with $q$. Equation (\ref{abar} differs with what in Purcell (1979) and LR97 by an additional term arising from variable transformation.

However, while considering grain dynamics, we introduce, in addition to the regular external torques, stochastic external torques. These torques are inseparable part of
the grain dynamics. Following Lazarian \& Draine (1997), we consider both stochastic torques arising from H$_2$ formation and those arising from gaseous bombardment. Our
treatment can be easily generalized for a different nature of the pinwheel torques.
The diffusion coefficients due to all active sites on the grain surface in the grain-body system $\hat{x},\hat{y}$ and $\hat{z}$, are given by

\bea
\langle (\Delta J_{x})^{2}\rangle=\gamma_{H} n_{H}a^{2} v_{H}a^{2}p_{0}^{2}\Gamma_{\|},\label{eq21}\\
\langle (\Delta J_{y})^{2}\rangle=\langle (\Delta J_{x})^{2}\rangle,\label{eq22}\\
\langle (\Delta J_{z})^{2}\rangle=\gamma_{H} n_{H}a^{2} v_{H}a^{2}p_{0}^{2}\Gamma_{\perp},\label{eq23}
\ena
where $\gamma_{H}$ is the fraction of $H$ atom converted to H$_{2}$, $n_{H}$ is the gas density, $v_{H}$ is the velocity of incoming atom, and $T_{gas}$ is the gas temperature, and $\Gamma_{\|},\Gamma_{\perp}$ are geometrical factors depending on the grain shape (see RDF93). For simplicity, we assume $\Gamma_{\|}=\Gamma_{\perp}=1$. In dimensionless units,
\bea
\langle (\Delta J_{x}')^{2}\rangle=\frac{E_{kin}}{4k_{B}T_{gas}},\label{eq21b}\\
\langle (\Delta J_{y}')^{2}\rangle=\langle (\Delta J_{x}')^{2}\rangle,\label{eq22b}\\
\langle (\Delta J_{z}')^{2}\rangle=\frac{E_{kin}}{4k_{B}T_{gas}},\label{eq23b}
\ena
where $E_{kin}=p_{0}^{2}/2m_{H}$ is the kinetic energy of evaporating $H_{2}$ molecule, and we neglected the anisotropy of diffusion coefficients. Here we adopt $E_{kin}=0.2 eV$ as suggested by Purcell (1979).
Transforming the diagonal tensor $B'_{ij}$ with $B'_{ii}=\langle (\Delta J_{ii}')^{2}\rangle$, and $B'_{i\ne j}=0$ from the grain body system to the spherical system described by $J,\theta, \phi$, we obtain the component along $\theta$ axis
\bea
   B'_{H2}({\theta})=0.5(B'_{11}+B'_{22})\left(\mbox{sin}^{4}\theta+0.5\mbox{cos}^{2}\theta+\mbox{sin}^{2}\theta\right)\nonumber\\
+1.5B'_{33}(\mbox{sin}^{2}\theta \mbox{cos}^{2}\theta),\label{bh2}
\ena
with $B'_{11}=B'_{22}=\langle (\Delta J'_{x})^{2}\rangle$, $B'_{33}=\langle (\Delta J'_{z})^{2}\rangle$.

The fluctuations arising from gas bombardment contribute also to the grain flipping. The diffusion coefficients obtained in Roberge et al. (1993), in the zeroth approximation, are given in the grain body system of reference by
\bea
B_{11}^{coll}&=&\frac{1}{4}n_{H}mv_{H}kT_{gas}b^{4}\left(1+(\kappa-1)\frac{T_{gas}}{T_{d}}\right)(\Lambda_{\perp}+\Gamma_{\perp}),\label{col1}\\
B_{22}^{gas}&=&B_{11}^{coll},\label{col2}\\
B_{33}^{coll}&=&\frac{1}{4}n_{H}mv_{H}kT_{gas}b^{4}\left(1+(\kappa-1)\frac{T_{gas}}{T_{d}}\right)(\Lambda_{\|}+\Gamma_{\|}),\label{col3}
\ena
where $n_{H}, v_{H}$ are density and velocity of atomic gas, respectively, $T_{gas}, T_{d}$ are temperature of gas and dust, $\kappa$ is the fraction of H atoms sticking to the grain that evaporate, $\Lambda_{\|}$ and $\Lambda_{\bot}$ are geometrical factors of order of unity.

In dimensionless units $B_{ij}^{'coll}=B_{ij}^{coll}J_{th}^{2}/t_{gas}$, we obtain
\bea
B_{11}^{'coll}&=&\left(1+(\kappa-1)\frac{T_{gas}}{T_{d}}\right)(\Lambda_{\perp}+\Gamma_{\perp}),\\
B_{22}^{'coll}&=&B_{11}^{'coll},\\
B_{33}^{'coll}&=&\left(1+(\kappa-1)\frac{T_{gas}}{T_{d}}\right)(\Lambda_{\|}+\Gamma_{\|})
\ena

\section{Flipping formalism}
Below we present the Fokker-Planck equation approach to study the flipping dynamics of grains. We note that this approach is presented in the classic book of Gardiner (1983), and brought to grain dynamics context by Roberge \& Ford (2000; RF00).\footnote{We provide more details of the RF00 approach,
as this paper has circulated only in the form of the preprint and may not be readily available. Their treatment is more rigorous than that in LD99a.}

Consider a brick grain with sides $b\times b\times a$ and $a<b$. The energy of this grain with temperature $T_{d}$ is given by
\bea
E(\theta)=\frac{J^{2}}{2I_{1}}\left( 1+[h-1]\ms^{2}\theta\right),\label{ap1}
\ena
where $I_{1}$ is the moment of inertia along the maximal inertia axis $\ba_{1}$, $T_{d}$ is the dust temperature, $\theta$ is the angle between $\ba_{1}$ and the angular momentum $\bJ$ of magnitude $J$, $k_{B}$ is the Boltzmann constant. 
Thermal fluctuations induce changes of $\theta$ and when $\theta>\pi/2$, the grain undergoes a flipping.

We assume that at time $t=0$, the grain is present at a given position $\theta\in[0,\pi/2]$. We can ask, after a time interval $t$, what is the probability that the grain is still present in the initial range. We denote such a probability by $G(\theta,t)$. After the time $t$, the grain can be present at position  $\theta'\in[0,\pi/2]$ with the probability $f(\theta',t)$, so that we have
\bea
G(\theta,t)=\int_{0}^{\pi/2}f(\theta',t)d\theta'.\label{gg}
\ena
We use the conditional probability $p(\theta_{2},t_{2}|\theta_{1},t_{1})$, i.e., the probability that the grain present at $\theta_{1},t_{1}$ will be present at $\theta_{2}$ at the time $t_{2}$. In terms of $p(\theta_{2},t_{2}|\theta_{1},t_{1})$
\bea
f(\theta',t)=p(\theta',t|\theta,0)=p(\theta',0|\theta,-t),
\ena
where the conditional probability which is uniform in time is adopted.

 Hence, the probability of flipping from the range $[0,\pi/2]$ to $[\pi/2,\pi]$ is
\bea
\Phi(\theta,t)=-\frac{G(\theta,t+\Delta t)-G(\theta,t)}{\Delta t}=-\frac{\partial G(\theta,t)}{\partial t}.\label{ap2}
\ena
To find $G(\theta, t)$ from equation (\ref{gg}), we need to know $f(\theta',t)$. The initial condition $t=0$ states that 
\bea
f(\theta',t=0)=\delta(\theta'-\theta).
\ena
Thus, 
\bea
G(\theta,t)=\int_{0}^{\pi/2}f(\theta',t)d\theta'=\int_{0}^{\pi/2}p(\theta',0|\theta,-t)d\theta'.\label{eq40}
\ena
It is well known that if $f$ satisfies the Fokker-Planck equation, then 
the probability $p$ with initial condition satisfies the backward Fokker-Planck equation:
\bea
\frac{\partial p(\theta',0|\theta,t)}{\partial t}=-A_{\theta}\frac{\partial p(\theta',0|\theta,t)}{\partial \theta}-\frac{B_{\theta\theta}}{2}\frac{\partial^{2}p(\theta',0|\theta,t)}{\partial\theta^{2}}.\label{eqa4a}
\ena
Integrating equation (\ref{eqa4a}) with $\theta'$ from $0$ to $\pi/2$ combined with equation (\ref{eq40}) and then changing variabel $t\rightarrow -t$, we obtain
\bea
\frac{\partial G(\theta,t)}{\partial t}=A_{\theta}\frac{\partial G(\theta,t)}{\partial \theta}+\frac{B_{\theta\theta}}{2}\frac{\partial^{2}G(\theta,t)}{\partial\theta^{2}}.\label{eqa4b}
\ena

Initial and boundary conditions for equation (\ref{eqa4b}) are given by
\bea
G(\theta,t=0)&=1 \mbox{ for } 0<\theta<\pi/2,\label{ap4a}\\
\left.\frac{\partial G}{\partial t}\right|_{t=0}&=0.\label{ap4b}
\ena
Also, a condition for the grain to flip immediately at $\theta=\pi/2$ is
\bea
G\left(\theta=\pi/2,t\right)=0.\label{ap4c}
\ena
This boundary condition corresponds to an absorbing boundary, i.e., the grain "escapes" at $\theta=\pi/2$.
The problem of a particle escape with initial and absorbing boundary conditions (\ref{ap4c}) is discussed in a great detail in Gardiner (2004; \S 5.2.7).

The mean mean flipping time is defined by means of the flipping probability (eq. \ref{ap2}):
\bea
t_{tf}(\theta)=\int_{0}^{\infty}t\Phi(t)dt=-\int_{0}^{\infty}t \frac{\partial G(\theta,t)}{\partial t} dt= -tG|_{0}^{\infty}+\int_{0}^{\infty}G(\theta,t) dt=\int_{0}^{\infty}G(\theta,t) dt,\label{ap7b}
\ena
where we used $G(\theta, t=\infty)=0$.
 By integrating equation \ref{eqa4b} over time and taking use of equation (\ref{ap7b}), we obtain

\bea
\int_{0}^{\infty}dt\frac{\partial G(\theta,t)}{\partial t}=G(\theta,\infty)-G(\theta,t=0)=A_{\theta}\int_{0}^{t}dt\frac{\partial G(\theta,t)}{\partial \theta}+\frac{B_{\theta\theta}}{2}\int_{0}^{t}dt\frac{\partial^{2}G(\theta,t)}{\partial\theta^{2}}.\label{eqa5a}
\ena
Hence
\bea
-1=A_{\theta}\frac{\partial \langle t_{tf}\rangle(\theta)}{\partial \theta}+\frac{B_{\theta\theta}}{2}\frac{\partial^{2}\langle t_{tf}\rangle(\theta)}{\partial\theta^{2}}.\label{eqa5b}
\ena
 The initial and boundary conditions for equation (\ref{eqa5b}) read
\bea
t_{tf}(\theta=\pi/2)=0,\label{eq42}\\
\frac{d t_{tf}}{d\theta} (\theta=0)=0.\label{eq43}
\ena
 The boundary condition (\ref{eq43}) implies that the grain flips immediately when the maximal inertia axis perpendicular to the angular momentum (i.e., $\pi/2$ is an absorbing boundary).

The solution of equation (\ref{eqa5b}) can be given by (Gardiner 1994; RF00)
\bea
t_{tf}(\theta)=\int_{\theta}^{\pi/2}\frac{d\theta'}{\Psi(\theta')}\int_{0}^{\theta'}\frac{2\Psi(\theta'')}{B_{\theta\theta}(\theta'')},\label{eqa7}
\ena
where 
\begin{equation}
\Psi(\theta')=\exp\left(\int_{\theta'}^{\theta}\frac{2A_{\theta}}{B_{\theta\theta}}d\theta''\right).\label{eqa8}
\end{equation}

\section{AMO and RATs components}
Here we present RATs produced by a radiation beam on a helical grain as shown in Figure \ref{amo}, and the averaging of RATs over thermal fluctuations.
\subsection{C1. RATs and mean values  over the spectrum of radiation fields}
The radiative torque resulting from the interaction of radiation field with a grain of size $a$ is defined by
\bea
{\bf \Gamma}_{rad}=\frac{\gamma u_{rad}\bar{\lambda} a^{2}}{2}\overline{\bf Q}_{\Gamma},\label{rat0}
\ena
where $\gamma$, $\bar{\lambda}$, and $u_{rad}$ are the degree of anisotropy, mean wavelength and total energy density of radiation field, respectively,  ${\bf Q}_{\Gamma}$ is the radiative torque efficiency vector, and overlines denote the averaging over the spectrum of the incident radiation field. 

Let $u_{\lambda}$ be the energy density per unit wavelength $\lambda$, the mean wavelength and total energy density for the radiation field is defined as
\bea
\overline{\lambda}=\frac{\int u_{\lambda} \lambda d\lambda}{\int u_{\lambda} d\lambda},\label{lamda}\\
{u}_{rad}=\int u_{\lambda} d\lambda.\label{urad}
\ena
Let $Q_{\Gamma}$ be the radiative torque efficiency produced by the monochromatic radiation field with wavelength $\lambda$, the mean torque efficiency over the radiation field is 
\bea
\overline{Q_{\Gamma}}=\frac{\int Q_{\Gamma}u_{\lambda} d\lambda}{\int u_{\lambda} d\lambda}\label{torqmean}.
\ena

\subsection{C2. General expressions of RATs}
RAT for the toy model in Figure \ref{amo} is given by the similar form as equation (\ref{rat0}):
\bea
{\bf \Gamma}_{rad}&=\frac{\gamma u_{rad}\bar{\lambda} l_{2}^{2}}{2}\overline{\bf Q}_{\Gamma},\label{a1}
\ena
where $l_{2}$, defined as the grain size $a$, is the size of the squared mirror.

Using the self-similar scaling of the magnitude of RATs obtained for irregular grains of size $a$ induced by the radiation field of wavelength $\lambda$,
\bea
\left|Q_{\Gamma}\right|&\sim& 0.4\left(\frac{\lambda}{a}\right)^{-3} \mbox{ for~$\lambda > 1.8 a$},\label{a2}\\
&\sim& 0.4 \mbox { for~$\lambda \le 1.8 a$},\label{a3}
\ena
and the functional forms of RATs from the AMO, we can write RAT components as following
\bea
Q_{e_{1}}(\Theta, \beta,\Phi= 0)&=&\frac{\left|Q_{\Gamma}\right|q^{max}}{\sqrt{(q^{max})^{2}+1}}\frac{q_{e_{1}}(\Theta, \beta, \Phi=0)}{q_{e_{1}}^{max}},\label{a4}\\
Q_{e_{2}}(\Theta, \beta, \Phi=0)&=&\frac{\left|Q_{\Gamma}\right|}{\sqrt{(q^{max})^{2}+1}}\frac{q_{e2}(\Theta, \beta, \Phi=0)}{q_{e_{2}}^{max}},\label{a5}\\
Q_{e_{3}}(\Theta, \beta, \Phi=0)&=&\frac{\left|Q_{\Gamma}\right|q^{max}}{\sqrt{(q^{max})^{2}+1}}\frac{q_{e3}(\Theta, \beta, \Phi=0)}{q_{e_{3}}^{max}},\label{a6}
\ena
where
\bea
q_{e_{1}}(\Theta, \beta, \Phi=0)&=&-\frac{4l_{1}}{\lambda}C\left(n_{1}n_{2}\frac{[3\mc^{2}\Theta-1]}{2}+\frac{n_{1}^{2}}{2}\mc\beta\ms2\Theta 
-\frac{n_{2}^{2}}{2}\mc\beta\ms2\Theta-\frac{n_{1}n_{2}}{2}\mc2\beta\right),\label{a7}\\
q_{e_{2}}(\Theta, \beta,\Phi= 0)&=&\frac{4l_{1}}{\lambda}C\left(n_{1}^{2}\mc\beta\mc^{2}\Theta-\frac{n_{1}n_{2}}{2}\mc^{2}\beta\ms2\Theta-\frac{n_{1}n_{2}}{2}\ms2\Theta +n_{2}^{2}\mc\beta\ms^{2}\Theta\right),\label{a8}\\
q_{e_{3}}(\Theta, \beta,\Phi= 0)&=&\frac{4l_{1}}{\lambda}C n_{1}\ms\beta \left[n_{1}\mc\Theta-n_{2}\mc\beta\ms\Theta\right]+\left(\frac{b}{l_{2}}\right)^{2}\frac{2e a}{\lambda}(s^{2}-1)K(\Theta)\ms 2\Theta,\label{a9}
\ena
with $q_{e_{j}}^{max}=max\langle q_{e_{j}}(\Theta, \beta,\Phi= 0)\rangle_{\beta}$ for $j=1,2$ and $3$. 
The magnitude ratio of torque components $q^{max}$ is defined by
\bea
q^{max}=\frac{max{\langle Q_{e_{1}}(\Theta, \beta, \Phi=0)\rangle_{\beta}}}{max{\langle Q_{e_{2}}(\Theta, \beta, \Phi=0)\rangle_{\beta}}}.\label{qmax}
\ena

In equations (\ref{a7})-(\ref{a9}), $C$ is a function given by
\bea
C=\left|n_{1}\mc\Theta-n_{2}\ms\Theta\mc\beta\right|,
\ena
where $\Theta$ is the angle between the axis of major inertia ${\bf a}_{1}$ and the radiation direction ${\bf k}$, $\beta$ is the angle describing the rotation of the grain about $\ba_{1}$ (see Fig. \ref{amo}{\it lower}); $n_{1}=-\ms i, n_{2}=\mc i$ are components of the normal vector
of the mirror tilted by an angle $i$ in the grain coordinate system, $a, b$ are minor and major semi-axes of the spheroid, $s=a/b<1$ and $e$ is the eccentricity of the spheroid, $l_{1}$ is the distance from the mirror to the spheroid, and $l_{2}$ is the size of the squared mirror; $K(\Theta)$ is the fitting function (see also LH07a). The second term of equation (\ref{a9}) represents the torque due to the spheroid. Assuming $l_{1}\sim \lambda$ and $b,~ a \sim l_{2} \ll l_{1}$, then this term is subdominant compared to the first term. Thus, we disregard it in our calculations (see Fig. \ref{amo}).

Our calculations for the alignment in the presence of thermal fluctuations  showed that the AMO can reproduce the alignment property with low-J as found with RATs obtained by DDSCAT when $q_{e_{1}}$ is modified to (see HL08a)
\bea
q_{e_{1}}(\Theta, \beta, \Phi=0)&=&-\frac{4l_{1}}{\lambda}C\left(n_{1}n_{2}\frac{[3\mc^{2}\Theta-1]}{2}+\frac{n_{1}^{2}}{2}\mc\beta\ms2\Theta 
-\frac{n_{2}^{2}}{2}\mc\beta\ms2\Theta-{n_{1}n_{2}}\mc2\beta\right).\label{a10}
\ena
This modification can arise from the imperfect scattering and/or the absorption effect by the mirror (LH07a). Also, the results in LH07a remain unchanged because the averaging over $\beta$ for the last term goes to zero.

We adopt the AMO with $i=45^{\circ}$ in this paper, unless mentioned otherwise. We also assume the amplitude of $Q_{e_{3}}$ is comparable to that of $Q_{e_{1}}$ and $Q_{e_{2}}$.

RATs at a precession angle $\Phi$ (see Fig. \ref{amo}{\it lower}) can be derived from RATs at $\Phi=0$ using the coordinate system transformation, as follows:
\bea
Q_{e_{1}}(\Theta, \beta, \Phi)&=&Q_{e_{1}}(\Theta, \beta, \Phi=0),\label{aeq4}\\ 
Q_{e_{2}}(\Theta, \beta, \Phi)&=&Q_{e_{2}}(\Theta, \beta,\Phi=
0)\mbox{cos}\Phi+Q_{e_{3}}(\Theta, \beta,\Phi=
0)\mbox{sin}\Phi,\label{aeq5} 
\\  
Q_{e_{3}}(\Theta, \beta, \Phi)&=&Q_{e2}(\Theta, \beta,\Phi=
0)\mbox{sin}\Phi-Q_{e_{3}}(\Theta, \beta,\Phi=
0)\mbox{cos}\Phi.\label{aeq6}
\ena

To study the alignment of the angular momentum with respect to magnetic field, we use the  spherical coordinate $J,\xi$ and $\phi$ (see Fig. \ref{sys}). In this coordinate system, RATs components are given by
\bea
F(\psi,\phi,\xi)&=&Q_{e_{1}}(\xi, \psi, \phi)(-\mbox{sin }\psi \mbox{cos }\xi \mbox{cos }\phi-\mbox{sin  }\xi \mbox{cos }\psi)+Q_{e_{2}}(\xi, \psi, \phi)(\mbox{cos }\psi \mbox{cos }\xi \mbox{cos }\phi-\mbox{sin }\xi
\mbox{sin }\psi)\nonumber\\
&&+Q_{e_{3}}(\xi, \psi, \phi)\mbox{cos }\xi \mbox{sin }\phi,\label{eeq9}\\ 
G(\psi,\phi,\xi)&=&Q_{e_{1}}(\xi, \psi, \phi)\mbox{sin }\psi \mbox{sin }\phi-Q_{e_{2}}(\xi, \psi, \phi)\mbox{cos }\psi \mbox{sin }\phi+Q_{e_{3}}(\xi, \psi, \phi)\mbox{cos }\phi,\label{eeq10}\\ 
H(\psi,\phi,\xi)&=&Q_{e_{1}}(\xi, \psi, \phi)(\mbox{cos }\psi \mbox{cos }\xi -\mbox{sin }\psi \mbox{sin  }\xi \mbox{cos }\phi)+Q_{e_{2}}(\xi, \psi, \phi)(\mbox{sin}\psi \mbox{cos }\xi + \mbox{cos }\psi\mbox{sin }\xi\mbox{cos  }\phi)\nonumber\\
&&+Q_{e_{3}}(\xi, \psi, \phi)\mbox{sin }\xi \mbox{sin }\phi, \label{eeq11} 
\ena
where $Q_{e_{1}}(\xi, \psi, \phi), Q_{e_{2}}(\xi, \psi, \phi), Q_{e_{3}}(\xi, \psi, \phi)$, as functions of $\xi, \psi$ and $\phi$, are components of the RAT efficiency vector in the
lab coordinate system (see DW97; LH07a). To obtain $Q_{e_{1}}(\xi, \psi, \phi),
Q_{e_{2}}(\xi, \psi, \phi)$ and $Q_{e_{3}}(\xi, \psi, \phi)$ from ${\bf
  Q}_{\Gamma}(\Theta, \beta, \Phi)$, we need to use the
relations between $\xi, \psi, \phi$ and $\Theta, \beta, \Phi$ (see WD03; HL08a).

\section{Flipping probability}

Once the mean flipping time $t_{tf}$ is known, one can calculate how much time the grain spends in each flipping state during a time interval $\Delta t$.

WD03 proposed an algorithm to calculate such a flipping probability. Let $P_{0}$ be the probability that the grain does not flip (i.e., $G(t)$ in Appendix B), and $f_{same}$ is the probability the grain stays in the same initial flipping state after $\Delta t$, they derived
\bea
f_{same}=\frac{1}{2}\left(1+\frac{exp(-\Delta t/t_{tf})sinh(\Delta t/t_{tf})}{\Delta t/t_{tf}}\right).\label{ap116}
\ena
The probability that the grain in the original flipping state after $\Delta t$ is the total probability that the grain does not flip plus the probability that the grain flips back original state if flip occurs
\bea
f_{same}=P_{0}+(1-P_{0})f_{1}+(1-P_{0})f_{2}...,
\ena
where $f_{i}$ is the fraction of time the grain spends in original state before the $i$ flipping occurs. Hence,
\bea
f_{same}=P_{0}+(1-P_{0})f_{s}
\ena
The fraction of time $\Delta t$ the grain spends in the initial flipping state is then 
\bea
f_{s}=\frac{f_{same}-P_{0}}{1-P_{0}},\label{ap114}
\ena
where 
\bea
P_{0}=exp(-\frac{\Delta t}{t_{tf}}).\label{ap115}
\ena


\begin{thebibliography}{8.}
\bibitem{} Aitken, D.K., Efstathiou, A., McCall, A.,
Hough, J.H, 2002, MNRAS, 329, 647-669
\bibitem[]{709} Andersson, B-G, Potter, S.B. 2007, ApJ, 665, 369
\bibitem[]{710} Bethell, T., Cherpunov, A., Lazarian, A., Kim, J. 2007, ApJ, 663, 1055
\bibitem[]{711} Cho, J., Lazarian, A. 2005, ApJ, 631, 361 
\bibitem[]{712} Cho, J.,\& Lazarian, A. 2007, ApJ, 669, 1085 (CL07)
\bibitem[]{713} Crutcher, R.M., Nutter, D.J., Ward-Thompson, D., \\
\& Kirk, J.M. 2004, ApJ, 600, 279
\bibitem[]{715} Davis, L. Vistas in Astronomy, Volume 1, 1955, 336
\bibitem[]{716} Davis, L., \& Greenstein J.L. 1951, ApJ, 114, 206 

\bibitem[]{717} Dolginov, A.Z. 1972, Ap\&SS, 16, 337
\bibitem[]{718} Dolginov, A.Z., \& Mytrophanov, I.G. 1976, Ap\&SS, 43, 291
\bibitem[]{719} Dolginov, A.Z., Silant'ev, N.A. 1976, Ap\&SS, 43, 337
\bibitem[]{720} Draine, B. 1996, in ASP Conf. Ser 97, Polarimetry of the Instellar Medium, ed. W.G. Roberge \& D.C.B. Whittet (San Francisco:ASP),16
\bibitem[]{720} Draine, B., \& Flatau, P.J. 2004, User guide for the Discrete Dipole Approximation Code DDSCAT 6.1,astro-ph/0409262
\bibitem{} Draine, B.T. \& Lazarian A. 1998, ApJ, 494, L19-L22
\bibitem[]{720} Draine, B., \& Lazarian, A. 1999, ApJ, 512, 740
\bibitem[]{721} Draine, B., \& Weingartner, J. 1996, ApJ, 470, 551 (DW96)
\bibitem[]{722} Draine, B.  \& Weingartner, J. 1997, ApJ, 480, 633 (DW97)
\bibitem[]{722} Dwek, E. et al. 2008, ApJ, 676, 1029
\bibitem[]{723} Falceta-Golcaves, D., Lazarian, A., \& Kowal, G. 2008, ApJ, 679, 537 
\bibitem[] Gardiner C.W. 1983, Handbook of Stochastic Method (Springer-Verlag)
\bibitem[]{724} Gold, T. 1952a,  Nature, 169, 322   
\bibitem[]{725} Gold, T. 1952b, MNRAS, 112, 215
\bibitem[]{726} Goodman, A., Jones, T., Lada, E., \& Myers, P. 1995, ApJ, 448, 748

\bibitem[]{728} Hall, J. 1949, Science, 109, 166
\bibitem[]{729} Harwit, M. 1970, Nature, 226, 61

\bibitem[]{731} Hildebrand, R., Davidson, J. A., Dotson, J.L,
Wovell C.D., \\
Novak, G.,\& Vaillancourt, J.E. 2000, PASP, 112, 1215
\bibitem[]{734} Hildebrand, R. 2002, in {\it Astrophysical Spectropolarimetry},\\
                                ed. by J. Trujillo-Bueno,
       F. Moreno-Insertis, \& F. Sanchez \\
(Cambridge, UK: Cambridge
       Univ. Press), p. 265
\bibitem[]{739} Hiltner, W. 1949, Science, 109, 165

\bibitem[]{740} Hoang, T, Lazarian, A. 2008, MNRAS, 388, 117 (HL08a)
\bibitem[]{740} Hoang, T, Lazarian, A. 2008, ApJ, submitted (HL08b)

\bibitem {} Hough, J.H., et al. 1989, MNRAS, 241, 71
\bibitem[]{742} Jones J.V., \& Spitzer, L. 1967, ApJ, 146,943
\bibitem[]{743} Kim S.-H., \& Martin, P. 1995, ApJ, 444, 293
\bibitem[]{744} Lai, S.-P., Crutcher, R.M., Girart, J.M., \& Rao, R. 2002, ApJ, 566,
925
\bibitem[]{746} Landau, L.D, \& Lifshitz, E. M. 1976, Mechanics (Oxford: Perganon)
\bibitem[]{747} Lazarian, A. 1994, MNRAS, 268, 713
\bibitem[]{748} Lazarian, A. 1995, ApJ, 453, 229
\bibitem[]{749} Lazarian, A. 1997a, ApJ, 483, 296
\bibitem[]{750} Lazarian, A. 1997b, MNRAS, 288, 609
\bibitem[]{755} Lazarian, A. 2003, J. Quant. Spectrosc. Rad. Trans., 79-80, 881
\bibitem[]{756} Lazarian, A. 2007, J. Quant. Spectrosc. Rad. Trans., 106, 225

\bibitem{} Lazarian, A., \& Draine, B.T., 1997, ApJ, 487, 248-258
\bibitem[]{757} Lazarian, A., \& Draine B. 1999a, ApJ, 516, L37 (LD99a)
\bibitem[]{758} Lazarian, A., \& Draine B. 1999b, ApJ, 520, L67 (LD99b)
\bibitem[]{752} Lazarian, A., \& Efroimsky, M. 1996, ApJ, 466, 274 
\bibitem[]{753} Lazarian, A., \& Efroimsky, M., Ozik J. 1996, ApJ, 472, 240
\bibitem[]{754} Lazarian, A., \& Efroimsky, M. 1999, MNRAS, 303, 673

\bibitem[]{759} Lazarian, A., \& Goodman A.A., Myers P.C. 1997, ApJ, 490, 273
\bibitem[]{760} Lazarian, A., \& Roberge W. 1997, ApJ, 484, 230 (LR97) 

\bibitem[]{762} Lazarian, A., \& Hoang T. 2007a, MNRAS, 378, 910 (LH07a)
\bibitem[]{763} Lazarian, A., \& Hoang T. 2007b, ApJL, 669, L77 (LH07b)

\bibitem[]{765} Lazarian, A., \& Yan H. 2002, ApJ, 566, L105-L108
\bibitem[]{766} Mathis, J. 1986, ApJ, 308, 281
\bibitem[]{767} Mathis, J., Mezger, P., \& Panagia, N. 1983, A\&A, 128, 212
\bibitem[]{331} Morrish A.H. 1980, The Physical Principles of Magnetism, Huntingdon: Krieger
\bibitem[]{768} Pagani, L. et al. 2004, A\&A, 413, 605
\bibitem[]{769} Pelkonen, V. M, Juvela, M., Padoan, P. 2007, A\&A, 461, 551
\bibitem[]{770} Purcell, E. 1969, Physica, 41, 100 
\bibitem[]{771} Purcell, E. 1975, in Dusty Universe, eds. Field G.B.,
Cameron, A.G.W., New York: Neal Watson, 155
\bibitem[]{773} Purcell, E. 1979, ApJ, 231, 404
\bibitem[]{774} Purcell, E., Spitzer, L. 1971, ApJ, 167, 31
\bibitem[]{775} Roberge, W., Hanany, S. 1990, B.A.A.S., 22, 862
\bibitem[]{776} Roberge, W., DeGraff, T.A., Flatherty, J.E. 1993, ApJ, 418, 287 (RDF93)
\bibitem[]{777} Roberge, W., Hanany, S., Messinger, D. 1995, 453, 238
\bibitem[]{778} Roberge, W.G., \& Lazarian, A. 1999, MNRAS, 305, 615
\bibitem[]{779} Roberge, W.G.,\& Ford, E., 2000, preprint (RF00)  
\bibitem{} Rosenbush, V.K., Kolokolova, L., Lazarian, A., Shakhovskoy, N.,
Kiselev, N. 2007, Icarus, 186, 317-330
\bibitem[]{780} Spitzer, L., McGlynn, T. 1979, ApJ, 231, 417
\bibitem[]{781} Spitzer, L, Tukey, 1951, ApJ, 114, 187

\bibitem[]{783} Ward-Thompson, D., Kirk, J.M, Crutcher, et al. 2000, ApJ, 537, L135
\bibitem[]{785} Ward-Thompson, D., Andre P.,  Kirk, J. 2002, MNRAS, 329, 257
\bibitem[]{786} Weingartner, J. \& Draine, B. 2003, ApJ, 589, 289
\bibitem[]{787_01} Weingartner, J. \& Jordan, M. 2008, ApJ, 672, 382 
\bibitem[]{787_02} Weingartner, J. 2008, ApJ, in press
\bibitem[]{788} Whittet, D.C.B., Gerakines, P.A., Hough, J.H.,\&  Shenoy 2001,
ApJ, 547, 872
\bibitem[]{790} Whittet, D.C.B, Hough, J.H, Lazarian, A., \& Hoang, T. 2008, ApJ, 674,304 
\end{thebibliography}
\end{document}